\documentclass[floatfix,pre,twocolumn,showpacs,superscriptaddress,notitlepage]{revtex4-1}

\usepackage{amsmath}
\usepackage{graphicx}
\usepackage{amssymb}
\usepackage{bm}
\usepackage{longtable}
\usepackage{color}

\usepackage[T1]{fontenc}
\usepackage{textcomp}
\usepackage{txfonts}
\usepackage{multirow}
\usepackage{hyperref}

\DeclareMathAlphabet{\mathitbf}{T1}{cmr}{bx}{it}

\begin{document}

\title{Dynamical transition in  the \boldmath $D =3$  Edwards-Anderson spin glass in an external magnetic field}

\author{M. Baity-Jesi} \affiliation{Departamento
  de F\'\i{}sica Te\'orica I, Universidad
  Complutense, 28040 Madrid, Spain.} \affiliation{Dipartimento di Fisica, La Sapienza Universit\`a di Roma,
     00185 Roma,  Italy.}\affiliation{Instituto de Biocomputaci\'on y
  F\'{\i}sica de Sistemas Complejos (BIFI), 50009 Zaragoza, Spain.}

\author{R.~A.~Ba\~nos} \affiliation{Instituto de Biocomputaci\'on y
  F\'{\i}sica de Sistemas Complejos (BIFI), 50009 Zaragoza, Spain.}
  \affiliation{Departamento
  de F\'\i{}sica Te\'orica, Universidad
  de Zaragoza, 50009 Zaragoza, Spain.} 

\author{A.~Cruz} \affiliation{Departamento
  de F\'\i{}sica Te\'orica, Universidad
  de Zaragoza, 50009 Zaragoza, Spain.} \affiliation{Instituto de Biocomputaci\'on y
  F\'{\i}sica de Sistemas Complejos (BIFI), 50009 Zaragoza, Spain.}

\author{L.A.~Fernandez} \affiliation{Departamento
  de F\'\i{}sica Te\'orica I, Universidad
  Complutense, 28040 Madrid, Spain.} \affiliation{Instituto de Biocomputaci\'on y
  F\'{\i}sica de Sistemas Complejos (BIFI), 50009 Zaragoza, Spain.}

\author{J.~M.~Gil-Narvion} \affiliation{Instituto de Biocomputaci\'on y
  F\'{\i}sica de Sistemas Complejos (BIFI), 50009 Zaragoza, Spain.}

\author{A.~Gordillo-Guerrero}\affiliation{D. de  Ingenier\'{\i}a
El\'ectrica, Electr\'onica y Autom\'atica, U. de Extremadura,
  10071, C\'aceres, Spain.}\affiliation{Instituto de Biocomputaci\'on y
  F\'{\i}sica de Sistemas Complejos (BIFI), 50009 Zaragoza, Spain.}

\author{D.~I\~niguez} \affiliation{Instituto de Biocomputaci\'on y F\'{\i}sica
  de Sistemas Complejos (BIFI), 50009 Zaragoza, Spain.}
\affiliation{Fundaci\'on ARAID, Diputaci\'on General de Arag\'on, Zaragoza, Spain.}

\author{A.~Maiorano} \affiliation{Dipartimento di Fisica, La Sapienza Universit\`a di Roma,
     00185 Roma,  Italy.}\affiliation{Instituto de Biocomputaci\'on y F\'{\i}sica de Sistemas
  Complejos (BIFI), 50009 Zaragoza, Spain.}

\author{F.~Mantovani} \affiliation{Dipartimento di Fisica e Scienze della
  Terra, Universit\`a di Ferrara, and INFN, Ferrara, Italy.}

\author{E.~Marinari}\affiliation{Dipartimento di Fisica, IPCF-CNR, UOS
Roma Kerberos and INFN, La Sapienza Universit\`a di Roma, 00185 Roma,  Italy.} 

\author{V.~Martin-Mayor} \affiliation{Departamento de F\'\i{}sica Te\'orica I,
  Universidad Complutense, 28040 Madrid, Spain.} \affiliation{Instituto de
  Biocomputaci\'on y F\'{\i}sica de Sistemas Complejos (BIFI), 50009 Zaragoza,
  Spain.}

\author{J.~Monforte-Garcia} \affiliation{Instituto de Biocomputaci\'on y
  F\'{\i}sica de Sistemas Complejos (BIFI), 50009 Zaragoza, Spain.}
  \affiliation{Departamento
  de F\'\i{}sica Te\'orica, Universidad
  de Zaragoza, 50009 Zaragoza, Spain.} 

\author{A.~Mu\~noz Sudupe} \affiliation{Departamento
  de F\'\i{}sica Te\'orica I, Universidad
  Complutense, 28040 Madrid, Spain.} 

\author{D.~Navarro} \affiliation{D.  de Ingenier\'{\i}a,
  Electr\'onica y Comunicaciones and I3A, U. de Zaragoza, 50018 Zaragoza, Spain.}

\author{G.~Parisi}\affiliation{Dipartimento di Fisica, IPCF-CNR, UOS
Roma Kerberos and INFN, La Sapienza Universit\`a di Roma, 00185 Roma,  Italy.}

\author{S.~Perez-Gaviro} \affiliation{Instituto de Biocomputaci\'on y
  F\'{\i}sica de Sistemas Complejos (BIFI), 50009 Zaragoza, Spain.}
\affiliation{Fundaci\'on ARAID, Diputaci\'on General de Arag\'on, Zaragoza, Spain.}

\author{M.~Pivanti} \affiliation{Dipartimento di Fisica e Scienze della
  Terra, Universit\`a di Ferrara, and INFN, Ferrara, Italy.} 

\author{F.~Ricci-Tersenghi}\affiliation{Dipartimento di Fisica, IPCF-CNR, UOS
Roma Kerberos and INFN, La Sapienza Universit\`a di Roma, 00185 Roma,  Italy.}  

\author{J.~J.~Ruiz-Lorenzo} \affiliation{Departamento de F\'{\i}sica,
  Universidad de Extremadura, 06071 Badajoz, Spain.}\affiliation{Instituto de
  Biocomputaci\'on y F\'{\i}sica de Sistemas Complejos (BIFI), 50009 Zaragoza,
  Spain.}

\author{S.F.~Schifano} \affiliation{Dipartimento di Matematica e Informatica, Universit\`a di Ferrara
and INFN, Ferrara, Italy}

\author{B.~Seoane}\affiliation{Dipartimento di Fisica, La Sapienza Universit\`a di Roma,
     00185 Roma,  Italy.} \affiliation{Instituto de
  Biocomputaci\'on y F\'{\i}sica de Sistemas Complejos (BIFI), 50009 Zaragoza,
  Spain.}

\author{A.~Tarancon} \affiliation{Departamento
  de F\'\i{}sica Te\'orica, Universidad
  de Zaragoza, 50009 Zaragoza, Spain.} \affiliation{Instituto de Biocomputaci\'on y
  F\'{\i}sica de Sistemas Complejos (BIFI), 50009 Zaragoza, Spain.}

\author{R.~Tripiccione} \affiliation{Dipartimento di Fisica e Scienze della
  Terra, Universit\`a di Ferrara, and INFN, Ferrara, Italy.} 
 
\author{D.~Yllanes}\affiliation{Dipartimento di Fisica, La Sapienza Universit\`a di Roma,
     00185 Roma,  Italy.}  \affiliation{Instituto de
  Biocomputaci\'on y F\'{\i}sica de Sistemas Complejos (BIFI), 50009 Zaragoza,
  Spain.}

\begin{abstract}

We study the off-equilibrium dynamics of the three-dimensional Ising spin glass
in the presence of an external magnetic field. We have performed simulations
both at fixed temperature and with an annealing protocol.  Thanks to the Janus
special-purpose computer, based on FPGAs,  we have been able to reach times
equivalent to $0.01$ seconds in experiments. We have studied the system
relaxation both for high and for low temperatures, clearly identifying a
dynamical transition point. This dynamical temperature is strictly positive and
depends on the external applied magnetic field. We discuss different
possibilities for the underlying physics, which include a thermodynamical
spin-glass transition, a mode-coupling crossover or an interpretation
reminiscent of the random first-order picture of structural glasses.
\end{abstract}

\maketitle 
\section{Introduction}

The glass transition is a ubiquitous but still mysterious phenomenon in
condensed matter physics \cite{debenedetti:97,debenedetti:01,cavagna:09}.
Indeed, many materials display a dramatic increase in their relaxation times when
cooled down to their glass temperature $T_\mathrm{g}$. Examples include
fragile molecular glasses, polymers, colloids or, more relevant here,
disordered magnetic alloys named spin glasses~\cite{mydosh:93}. The dynamic
slowing down is not accompanied by dramatic changes in structural or
thermodynamic properties. In spite of this, quite general arguments suggest
that the sluggish dynamics must be correlated with an increasing length
scale~\cite{montanari:06}. This putative length scale can be fairly difficult
to identify (a significant amount of research has been devoted to this
problem, see, e.g.,~\cite{adam:65,weeks:00,berthier:05}).

In this context, spin glasses are unique for a number of reasons. To start
with, they provide the only example of a material for which it is widely
accepted that the sluggish dynamics is due to a thermodynamic phase transition
at a critical temperature
$T_\text{c}=T_\mathrm{g}$~\cite{gunnarsson:91,palassini:99,ballesteros:00}.
This phase transition is continuous, and the time-reversal symmetry of the
system is spontaneously broken at $T_\text{c}$. 

Spin glasses are also remarkable because special tools are available for their
investigation. On the experimental side, time-dependent magnetic fields are
a very flexible tool to probe their dynamic response, which can be very
accurately measured with a SQUID (for instance, see~\cite{herisson:02}).
For very low applied magnetic fields, one can
measure {\em glassy} magnetic domains of a diameter up to $\xi \sim 100$
lattice spacings~\cite{joh:99,bert:04}, much larger than any length scale
identified for structural glasses~\cite{berthier:05}. On the theoretical side,
spin glasses are simple to model, which greatly eases numerical simulation. In
fact, special-purpose computers have been built for the simulation of spin
glasses~\cite{cruz:01,ogielski:85,janus:06,janus:08, janus:08b}. In particular,
the Janus computer~\cite{janus:06,janus:08} has allowed us to simulate the
non-equilibrium dynamics from picoseconds to a tenth of a
second~\cite{janus:08b,janus:09b}, which has resulted in detailed connections
between non-equilibrium dynamics at finite times and equilibrium physics in
finite systems \cite{janus:10,janus:10b} (see also \cite{franz:98}).

Now, Mean Field provides compelling motivation to investigate spin glasses in
an externally applied magnetic field. Although a magnetic field explicitly
breaks time-reversal symmetry, it has been shown that a phase transition
should still occur when cooling the mean-field spin glass in an external
field~\cite{dealmeida:78,mezard:87,marinari:00}, which leads us to expect
a sophisticated and still largely unexplored behavior for short-range
spin glasses. In this respect, we
remark as well an intriguing suggestion by Moore, Drossel and Fullerton.
These authors speculate that the spin glass in a field sets the universality
class for the structural-glass transition \cite{moore:02,fullerton:13} (the statement is supposed to hold
in $d=3$ spatial dimensions; spatial dimensionality \emph{is}
crucial).

However, two major problems hamper further progress: (i) the mean-field
theory, which is expected to be accurate above the upper critical dimension
$d>d_\text{u}=6$, predicts a rather different behavior for spin and structural
glasses; (ii) the behavior of spin glasses in a field in $d=3$ is the matter
of a lively controversy. We now address each issue separately.

Regarding the first problem, we note that the standard mean-field model for
structural glasses is the $p$-spin glass model, with $p$-body interactions 
\cite{kirkpatrick:87,kirkpatrick:89}
(for odd $p$, the time-reversal symmetry is explicitly
broken). In the mean-field approximation,
the odd-$p$ models display a dynamic phase transition~\emph{in their
  paramagnetic phase.\/} Reaching thermal equilibrium becomes impossible in
the temperature range $T_{\mathrm c}<T<T_\text{g}$. The dynamic transition at
$T_\text{g}$ is identical to the ideal Mode Coupling transition of supercooled
liquids~\cite{goetze:92}. The thermodynamic phase transition at $T_{\mathrm c}$
is analogous to the ideal Kauzmann thermodynamic glass
transition~\cite{cavagna:09}. The thermodynamic transition is very peculiar:
although it is of the second order (in the Ehrenfest sense), the spin-glass
order parameter jumps discontinuously at $T_\mathrm{c}$ from zero to a
non-vanishing value. On the other hand, for spin glasses in a field,
mean-field theory predicts a single transition (i.e., $T_{\mathrm
  c}=T_\text{g}$) and an order parameter that behaves continuously as a
function of temperature.

However, mean-field theory is accurate only in high-enough spatial dimensions, 
$d>d_\text{u}=6$, hence it is legitimate to wonder about our three-dimensional
world. In fact, the ideal Mode Coupling transition is known to be only a
crossover for supercooled liquids.  The power-law divergences
predicted by Mode Coupling theory hold when the equilibration time lies in the
range $10^{-13}\text{ s} <\tau < 10^{-5} \text{ s}$.  Fitting to those
power-laws, one obtains a Mode Coupling temperature
$T_{\mathrm{MC}}$. However, $\tau$ is finite at $T_{\mathrm{MC}}$ (typically
$T_{\mathrm{MC}}$ is 10\% larger than the glass temperature $T_\text{g}$ where
$\tau\sim 10^4$ seconds). A theory for a thermodynamic glass transition at
$T_\mathrm{c}<T_\text{g}$ has been put
forward~\cite{mezard:99,mezard:99b,coluzzi:00,parisi:10}, but it has still not
been validated (however, see~\cite{singh:13}).

Regarding now our second problem, we recall that whether spin glasses in a
magnetic field undergo a phase transition has been a long-debated and still
open question. There are mainly two conflicting theories. The above mentioned
mean-field picture is the replica symmetry breaking (RSB)
theory~\cite{mezard:87}. On the other hand, the droplet
theory~\cite{mcmillan:84,bray:87,fisher:86,fisher:88b} predicts that the
system's behavior is akin to that of a disguised ferromagnet (i.e., no phase
transition in a field). Recent exponents of this controversy are
\cite{moore:11,parisi:12,yeo:12}. We elaborate on concrete
predictions by both theories in Section~\ref{sec:droplets_vs_RSB}. For now, we
note that recent numerical simulations in $d=3$ did not find the thermodynamic
transition predicted by mean-field~\cite{young:04,jorg:08b}. Experimental
studies have been conducted as well, with conflicting
conclusions~\cite{jonsson:05,petit:99,petit:02,tabata:10}. Up to now, a
transition has been found only numerically, in four dimensions (note that
$4<d_\mathrm{u}=6$)~\cite{janus:12}.

A different numerical approach has consisted in the study of Levy graphs:
one-dimensional systems where the interaction decays with distance as a
power-law $J(r) \sim 1/r^{\hat a}$~\cite{kotliar:85}. It has been suggested
that the critical behavior in $d$ spatial dimensions can be matched with that
of a Levy graph with an appropriate choice of the decay exponent $\hat a$. The
spin-glass transition has been investigated in this way, both with and without
an externally applied magnetic field. The latest result within this approach
suggests that the de Almeida-Thouless line is present in four spatial
dimensions, but not in three dimensions~\cite{larson:13}. However, it has been
pointed out that the matching between the decay exponent $\hat a$ and the
spatial dimension $d$ changes when a magnetic field is
applied~\cite{leuzzi:13}, a possibility not considered in
\cite{larson:13}.

Our scope here is to explore the dynamical behavior of $d=3$ spin glasses in a
field using the Janus computer. We shall study lattices of size $L=80$, where
we expect finite-size effects to be negligible \cite{janus:08b}. Our time
scales will range from 1 picosecond (i.e., one Monte Carlo full lattice
sweep~\cite{mydosh:93}) to $0.01$ seconds.  Hence, if the analogy with
structural glasses holds, we should be able to identify the Mode Coupling
crossover. Our study will be eased by the rather deep theoretical knowledge of
the relevant correlation functions \cite{dedominicis:06}. Hence, we shall be
able to correlate the equilibration time $\tau$ with the correlation length
$\xi$.

The layout of the remaining part of this work is as follows.  In
Section~\ref{sec:MODELand_observables} we describe the model and our
observables and we give an overview of the different theoretical pictures that
have been put forward to explain the dynamics of the spin glass in a field. We
pay particular attention to the specific predictions for the observables that
we are going to study (Sections~\ref{sec:droplets_vs_RSB} and
Section~\ref{sec:MCT-and-beyond}).  In Section~\ref{sec:PROTOCOLS_AND_TAUS} we
describe the different protocols that we have considered (simulating a direct
quench and annealings with different temperature variation rates).  Next, in
Section~\ref{sec:TAUS}, we turn our attention to the crucial and delicate
issue of identifying intrinsic time scales in the dynamics.

Once this foundation has been laid, we delve into the physical analysis and
interpretation of our results. First, we consider the high-temperature regime,
where our simulations reach equilibrium, in Section~\ref{sec:equilibrium}.  We
thus try to approach the critical region from above.  Afterwards, in
Section~\ref{sec:NON-EQUILIBRIUM}, we study the low-temperature regime, where
the relaxation times are much longer than our simulations (perhaps
infinite). We consider two complementary approaches: in
Section~\ref{sec:NON-EQUILIBRIUM-POWER-LAW} we try the simplest ansatz for a
low-temperature behavior compatible with a RSB spin-glass transition, while in
Section~\ref{sec:NON-EQUILIBRIUM-LINEAR-LOG} we assume from the outset that no
transition occurs. The spatial correlation of the system is studied in
Section~\ref{sec:xi}, where we carry out a direct analysis of the correlation
length.  Finally, in Section~\ref{sec:conclusions} we present our conclusions,
evaluating the different physical scenarios on the light of our data.

We also include two appendices: one describing some details of our
implementation and the other presenting a more detailed look at the
correlation functions.

%%%%MODEL
%%%%%%%%%%%%%%%%%%%%%%%%%%%%%%%%%%%%%%%%%%%%%%%%%%%%%%%%%%%%%%%%%%
\section{Model, observables and the droplet-RSB controversy}\label{sec:MODELand_observables}
%%%%%%%%%%%%%%%%%%%%%%%%%%%%%%%%%%%%%%%%%%%%%%%%%%%%%%%%%%%%%%%%%%

In Section~\ref{sec:MODEL} we describe the model that we have simulated.  The
observables that we consider are defined in Section~\ref{sec:OBSERVABLES}. At
that point, we will be ready to describe the different predictions of the
droplet and RSB theories in Section~\ref{sec:droplets_vs_RSB}. However, it has
been recently suggested that spin-glass on a field behave just as supercooled
liquids \cite{moore:02,fullerton:13}. The current theoretical predictions for
supercooled liquids do not quite match neither the droplet nor the RSB
theories for spin glasses on a field. Hence, we briefly recall those
predictions in Section~\ref{sec:MCT-and-beyond}.

\subsection{Model}\label{sec:MODEL}

We have studied a three-dimensional cubic lattice system with volume $V=L^3$ ($L$
is the linear size) and periodic boundary conditions. On every node of the
lattice there is an Ising spin $\sigma_\mathitbf{x}$ and nearest neighbors
are joined by couplings $J_\mathitbf{x y}$.  The spins are dynamic variables,
while the couplings $J_\mathitbf{x y}$ are fixed during the simulation (this is the
so-called quenched disorder \cite{parisi:94}). The couplings are
independent random variables: $J_\mathitbf{x y}=\pm 1$ with $50\%$
probability. We also include a local magnetic field $h_\mathitbf{x}$ on
every node. The magnetic field is Gaussianly distributed with zero mean and
variance $H^2$.  The Hamiltonian of the model is
\begin{equation}
  \label{eq:ham}
  {\cal H} = - \sum_{\langle \mathitbf{x},\mathitbf{y} \rangle}J_{\mathitbf{x y}} \sigma_\mathitbf{x} \sigma_\mathitbf{y} 
              - \sum_\mathitbf{x} h_\mathitbf{x} \sigma_\mathitbf{x},
\end{equation}
where $\langle \mathitbf{x,y} \rangle$ means sum over nearest neighbors. A
given realization of couplings, $J_\mathitbf{x y}$, and an external field,
$h_\mathitbf{x}$, defines a {\it sample}. All of our results will be averaged
over many samples.

It is important to realize that both the Hamiltonian~\eqref{eq:ham} and the
probability distribution functions (pdf) for the couplings and the magnetic
field are invariant under a gauge symmetry. Let be $\epsilon_\mathitbf{x}=\pm
1$ be arbitrary site-dependent signs. The gauge symmetry is
\begin{equation}\label{eq:gauge-symm}
\sigma_\mathitbf{x}\rightarrow \epsilon_\mathitbf{x} \sigma_\mathitbf{x}\,,\quad 
h_\mathitbf{x}\rightarrow \epsilon_\mathitbf{x} h_\mathitbf{x}\,,\quad 
J_\mathitbf{x y}\rightarrow \epsilon_\mathitbf{x} \epsilon_\mathitbf{y} 
J_\mathitbf{x y}\,.
\end{equation}
The observables defined in Section~\ref{sec:OBSERVABLES} must be invariant under
this symmetry. As explained in the literature (see,
e.g., \cite{janus:09b,janus:10}), one forms gauge-invariant observables by
considering real replicas, namely copies of the system that evolve
independently under the same coupling constants and magnetic fields. In
particular, we consider four replicas per sample.

The model~\eqref{eq:ham} has been simulated on the Janus special-purpose
computer \cite{janus:06,janus:08,janus:09,janus:12b}. Janus' limited memory 
does not allow us to study one independent, real-valued magnetic field
per site $h_\mathitbf{x}$. We circumvented the problem using Gauss-Hermite
integration \cite{abramowitz:72}. Details of our implementation, as well as
comparison with PC simulations for real-valued $h_\mathitbf{x}$, can be found
in Appendix~\ref{app1}.

From the theoretical point of view, two main and contradictory theoretical
predictions for the finite-dimensional model in an external magnetic field can
be found in the literature: the droplet theory (DT) and the Replica Symmetry
Breaking theory (RSB).

In the droplet framework, the $H=0$ spin-glass phase is destroyed by any
magnetic field, however small: there is only a paramagnetic
phase~\cite{fisher:86,bray:87,fisher:88b}.

Instead, in the RSB construction, the spin-glass phase remains in the presence
of small magnetic fields. The boundary of this spin-glass phase with the
paramagnetic phase is called the de Almeida-Thouless (dAT)
line~\cite{mezard:87,marinari:00}.

Finally, we can summarize briefly the main analytical results obtained up to
date.  Renormalization Group (RG) studies, assuming that the longitudinal and
anomalous sectors are degenerate  (see below for more details) have failed to
find a fixed point~\cite{bray:80}. The upper critical dimension in this
approach turns out to be just six as in absence of a magnetic field (however,
some observables change their mean-field behavior just at and below eight
dimensions due to the appearance of dangerous irrelevant variables in the
RG~\cite{fisher:85}).

The lack of stable fixed points below six dimensions could be due to the
absence of a phase transition, the presence of a first-order phase transition
or the limitations of the perturbative approach used (e.g., there might
be a stable fixed point for higher orders of the perturbative expansion). These
computations~\cite{bray:80} have been performed enforcing that the number of
replicas, $n$, of the effective field theory be zero, which implies the above
cited degeneration between the anomalous and longitudinal  sectors of the
theory.

However, \cite{temesvari:02} started using the most general
Hamiltonian compatible with the symmetries of a replica symmetric phase and
relaxed the $n=0$ condition, so the longitudinal and anomalous sectors are no
longer degenerate.  In this work stable fixed points were found below
six dimensions.  In addition, in a more recent work \cite{temesvari:08}, the
AT line was computed in dimensions slightly below the upper critical dimension
($d \lesssim 6$) (see also \cite{parisi:12}).  Notice that in all
these analytical studies the phase transition (if it exists) is approached
from the paramagnetic phase.  Hence, the only information about the structure
of a tentative spin-glass phase in finite dimensions is just that of mean field.

\subsection{Observables}\label{sec:OBSERVABLES}
%----------------------------------------------
We are going to consider the temporal evolution of different physical quantities
in simulation protocols with temperature variation.
To this end, we define $t_\text{tot}$ as the total time elapsed during the simulation
(measured in Monte Carlo Sweeps, MCS, i.e., updates of the whole lattice) and
the waiting time $t_\text{w}$ as the total time elapsed since the last 
change of temperature. When this distinction is not important, we use $t$ as a 
generic time variable.

First, a couple of useful definitions of local quantities.  On every node $\boldsymbol x$
of the lattice we have the local overlap:
\begin{equation}
 \label{eq:qx}
 q_\mathitbf{x}(t)=\sigma_\mathitbf{x}^{(1)}(t) \sigma_\mathitbf{x}^{(2)}(t)\,,
\end{equation}
where the superscripts are the replica indices.  The total overlap is written as
\begin{equation}
\label{eq:q}
 q(t)=\frac{1}{V}\overline{ \sum_\mathitbf{x} q_\mathitbf{x}(t)}\, ,
 \end{equation}
where $\overline{(\cdots)}$ means sample average (over the $J$'s and $h$'s),
and $V=L^3$ is the total number of spins in the lattice.

In addition, we
have focused in this work on the magnetic energy defined as
\begin{equation}
\label{eq:Emag}
E_\mathrm{mag}(t)=\frac{1}{V} \overline{ \sum_\mathitbf{x} h_\mathitbf{x} \sigma_\mathitbf{x}(t)} \, .
\end{equation}
and
\begin{equation}
\label{eq:W}
W(t)= 1-T E_\mathrm{mag} (t)/H^2 \, .
\end{equation}
We note that if the magnetic field is Gaussian distributed,
the following identity is fulfilled in \emph{thermal equilibrium}:
\begin{equation}
\label{eq:W_term}
W= \overline{\langle q
  \rangle}\,.
\end{equation}
To derive Eq.~\eqref{eq:W_term} integrate by parts the term $h_\mathitbf{x}
e^{-h_\mathitbf{x}^2/(2 H^2)}$ that appears on the disorder average for
$E_\mathrm{mag}$, and recall the \emph{equilibrium} fluctuation-dissipation
relation $\mathrm{d}\langle s_\mathitbf{x}\rangle/\mathrm{d}h_\mathitbf{x}=(1-\langle s_\mathitbf{x}\rangle^2)/T\/$.

Both $q(t)$ and $W(t)$ disregard the spatial dependence. To address this
issue, we should consider correlation functions. In a magnetic field, one may
consider three different correlators $G_1$, $G_2$ and $G_3$ at equal time. The
three of them can be computed with four replicas:
\begin{align}
G_1(\mathitbf{r},t)&=\frac{1}{V}\sum_{\mathitbf{x}}\,\overline{\langle
  \sigma_{\mathitbf{x}}(t) \sigma_{\mathitbf{x+r}}(t) \rangle^2} \nonumber\\
&= \frac{1}{V}\sum_{\mathitbf{x}}\,\overline{\langle
  \sigma_{\mathitbf{x}}^{(1)}(t) \sigma_{\mathitbf{x+r}}^{(1)}(t) \sigma_{\mathitbf{x}}^{(2)}(t) \sigma_{\mathitbf{x+r}}^{(2)}(t) \rangle}
\,,\\
G_2(\mathitbf{r},t)&=\frac{1}{V}\sum_{\mathitbf{x}}\,\overline{\langle \sigma_{\mathitbf{x}}(t) \sigma_{\mathitbf{x+r}}(t) \rangle\langle \sigma_{\mathitbf{x}}(t)\rangle \langle 
\sigma_{\mathitbf{x+r}}(t)\rangle} \nonumber\\
&= \frac{1}{V}\sum_{\mathitbf{x}}\,\overline{\langle
  \sigma_{\mathitbf{x}}^{(1)}(t) \sigma_{\mathitbf{x+r}}^{(1)}(t) \sigma_{\mathitbf{x}}^{(2)}(t) \sigma_{\mathitbf{x+r}}^{(3)}(t) \rangle}\,,\\
G_3(\mathitbf{r},t)&=\frac{1}{V}\sum_{\mathitbf{x}}\,\overline{\langle
  \sigma_{\mathitbf{x}}(t)  \rangle^2\langle \sigma_{\mathitbf{x+r}}(t)
  \rangle^2}\\
&= \frac{1}{V}\sum_{\mathitbf{x}}\,\overline{\langle
  \sigma_{\mathitbf{x}}^{(1)}(t) \sigma_{\mathitbf{x+r}}^{(2)}(t) \sigma_{\mathitbf{x}}^{(3)}(t) \sigma_{\mathitbf{x+r}}^{(4)}(t) \rangle}\,.\nonumber
\end{align}
To gain statistics, we average over all the replica-index permutations that
yield the same expectation values in the above equations. 

Now, because of the magnetic field, none of the correlators $G_1$, $G_2$ and
$G_3$ tend to zero for large $\mathitbf{r}$, hence one needs to define
\emph{connected} correlators.  This problem was faced long ago, and the answer
is in the three basic propagator of the replicated field theory namely the
longitudinal ($G_\text{L}$), anomalous ($G_\text{A}$) and replicon
($G_\text{R}$) (see \cite{dealmeida:78,bray:80}):
\begin{eqnarray}
G_\text{R}&=&G_1-2 G_2+  G_3\,,\label{eq:GR}\\
G_\text{L}&=&G_1+ 2 (n-2) G_2+ \frac{1}{2}(n-2)(n-3) G_3\,,\label{eq:GL} \\
G_\text{A}&=&G_1+(n-4) G_2-(n-3) G_3\,,\label{eq:GA}
\end{eqnarray}
where $n$ is the number of replicas of the effective replica field theory
($n$ should not be confused with the number of \emph{real replicas} that we
hold fixed to four). Quenched disorder is recovered from replica field theory
in the limit $n\to 0$. In this limit,
Eqs.~(\ref{eq:GR},~\ref{eq:GL},~\ref{eq:GA}) yield the replicon
\begin{equation}
\label{eq:C_r_tw}
G_\text{R}(\mathitbf{r}, t) = \frac{1}{V}
\sum_{\mathitbf{x}} \overline{\left( \langle \sigma_{\mathitbf{x}}(t) 
\sigma_{\mathitbf{x}+\mathitbf{r}}(t) \rangle 
- \langle \sigma_{\mathitbf{x}}(t) \rangle \langle 
\sigma_{\mathitbf{x}+\mathitbf{r}}(t) \rangle \right)^2}\, ,
\end{equation}
while $G_\text{A}$ and $G_\text{L}$ coalesce in the limit to a single propagator:
\begin{equation}\label{eq:GLGA_n_to_zero}
G_\text{L}=G_\text{A}=G_1-4 G_2+3 G_3\,,
\end{equation}
(strictly speaking, replica field theory applies only to equilibrium; the
reader will forgive us for borrowing the natural equilibrium definitions for
our dynamic computation at finite $t$).

Having in our hands two correlation functions (replicon and
longitudinal/anomalous), it is natural to compute the associated
susceptibilities
\begin{equation}\label{eq:chi_def}
\chi(t)=\int \text{d}^3\boldsymbol r ~C(\boldsymbol r,t)\,,
\end{equation}
where $C$ stands for any of the correlators introduced in this section.

At finite $t_\text{w}$, correlations surely are sizeable only within some
characteristic correlation length $\xi(t_\text{w})$. If one expects a typical
scaling form for such correlators
\begin{equation}
C(\mathitbf{r})\sim \frac{f(\mathitbf{r}/\xi(t_\text{w}))}{r^a}\,,
\end{equation}
where $f$ is some long-distance cutoff function, it is natural to extract the
correlation length from integral estimators~\cite{janus:08b,janus:09b}
    \begin{equation}
	\label{eq:xi_integrals}
       \xi_{k,k+1}(t_\text{w}) \equiv \frac{I_{k+1}(t_\text{w})}{I_k(t_\text{w})} \propto \xi(t_\text{w}) \, .
      \end{equation}
where
\begin{equation}
\label{eq:integrals}
       I_k(t_\text{w}) \equiv \int_0^{L/2}\text{d}r ~ r^kC(r,t_\text{w}) \,.
\end{equation}
The $r$ in Eq.~\eqref{eq:integrals} is a shorthand for $\mathitbf{r}=\left(
r,0,0 \right)$ and permutations. The signal-to-noise ratio is increased if one
imposes a long-distance cutoff in Eq.~\eqref{eq:integrals}. In particular, we
stop integrating when the relative error in $C(r,t_\text{w})$ grows larger
than $1/3$ (we can estimate the effect of the tail with an exponential
extrapolation, although this is a very minor effect, see
\cite{janus:09b} and Appendix~\ref{app_GRGL:truncacion}).

Notice that a similar method can be used to obtain estimators for the susceptibility.
Assuming isotropy (see \cite{janus:09b}) we can write Eq.~\eqref{eq:chi_def} as 
\begin{equation}
\chi(t) = 4 \pi \int_0^{L/2} \text{d} r\ r^2 C(r,t) = 4\pi I_2(t),
\end{equation}
which can be computed with the long-distance cutoff.

Finally, let us mention that quantities such as $C(r,t_\text{w})$, due to
the slow temporal evolution and the strong sample-to-sample fluctuations, 
turn out to be very rugged functions of $t_\text{w}$. This is particularly 
bad for the computation of $\xi_{12}$, since it can have an unpredictable
effect on the cutoff. In order to avoid this problem, we perform a temporal
binning, averaging $C(r,t_\text{w})$ in blocks of $4$ consecutive times
and considering it as a function of the geometrical mean of the times
in each block. This smoothing procedure yields a significant error reduction.

\subsection{The droplet versus RSB controversy in terms of our observables}\label{sec:droplets_vs_RSB}

We summarize here the major differences among the predictions from the droplet
and RSB theories with an emphasis on the quantities defined in
Section~\ref{sec:OBSERVABLES}. We discuss as well the different
predictions for the characteristic time scale for equilibration, $\tau$ (see
Section~\ref{sec:TAUS}). As we shall see, the predictions of the droplet theory
are more detailed, hence let us start there.

\subsubsection{Predictions from the droplet theory}

According to the droplet theory, there is no phase transition in a field.
This means that the long-time and large-system limits, $t_\text{w}\to\infty$
and $L\to \infty$, can be taken in any order. In this work, we will always
take \emph{first} the limit of large $L$ (because $L\gg \xi(t_\text{w})$ in
our simulations, see \cite{janus:08b}). Thus, it is a consequence
of the droplet picture that Eq.~\eqref{eq:W_term} should always hold for large
lattices and then large times.

The physics at low temperatures is predicted to be ruled by a fixed point
at $T=H=0$ \cite{bray:87,fisher:88b}. Temperature is an irrelevant scaling field, while the magnetic
field is relevant. The scaling dimensions would be
\begin{eqnarray}
y_T&=&-\theta\,,\label{ea:droplet_yT}\\
y_H&=&\frac{d- 2\theta}{2}\,,\label{ea:droplet_yH}
\end{eqnarray}
where $\theta$ is the droplet stiffness exponent and $d$ is the space
dimension ($d=3$ in this work). Therefore, the correlation length is predicted
to remain finite for all positive temperatures, even in the limit
$t_\text{w}\to\infty$. A scaling law follows from
Eqs.~(\ref{ea:droplet_yT},\ref{ea:droplet_yH}), which should hold at least for
small temperatures and magnetic fields:
\begin{equation}\label{eq:droplets-for-xi}
\xi(T,H)=\frac{1}{H^{2/(d-2\theta)}} F(T H^{2\theta/(d-2\theta)})\,.
\end{equation}
$F(x)$ is a scaling function, which is supposed to remain bounded when
$x\to 0$. 

As for the dynamics, the droplet prediction is that it is of the activated
type, with a typical barrier of order $\xi^\varPsi$ [$\varPsi$ is a second
droplet exponent which should satisfy $\theta \le \varPsi \le
(d-1)$~\cite{fisher:88b}]. Hence, since $\xi$ is expected to remain bounded at
all temperatures, the dynamics is predicted to be of Arrhenius type
\begin{equation}\label{eq:droplets-dynamics}
\tau \propto \mathrm{exp}\, \bigg[\frac{\xi^\varPsi(T,H)}{T}\bigg]\,.
\end{equation}
Of course, if $\xi$ grows significantly with $T$ for some temperature
interval, Eq.~\eqref{eq:droplets-dynamics} would predict an apparent
super-Arrhenius behavior.

Now, in order to make some use of
Eqs.~(\ref{eq:droplets-for-xi},\ref{eq:droplets-dynamics}), it is mandatory to
have an estimate for the exponents $\theta$ and $\varPsi$. For the barrier exponent
$\varPsi$, we may quote an experimental determination on the Ising system
Fe$_{0.5}$Mn$_{0.5}$TiO$_3$, $\varPsi\sim 0.03$ \cite{bert:04}. As for the
$\theta$ exponent, we are aware of two types of computations.  On the one
hand, the properties of ground states have been analyzed. The comparison of
periodic and antiperiodic boundary conditions yields
$\theta\sim0.2$ \cite{hartmann:99c,palassini:99b},
the most recent value being $\theta=0.24(1)$ \cite{boettcher:04}.
However, excitations of much lower energy, low enough in fact as to suggest
$\theta=0$, have been identified with fixed boundary
conditions \cite{palassini:00,krzakala:00}. On the other hand, one may address
the computation of $\theta$ in a rather more direct way, by considering the
behavior of the spatial correlation functions. This approach yields
$\theta=0.61(8)$ [the error estimate contains both statistical and systematic
  effects, see Eq. (11.64) in \cite{yllanes:11} and the related
  discussion]. Given this disparate range for the estimations of exponent
$\theta$, we shall use both $\theta=0.24$ and $\theta=0.61$ when
trying to assess Eq.~\eqref{eq:droplets-for-xi}.

\subsubsection{Predictions from the RSB theory}

The RSB theory is based on the solution of a mean-field model (the
Sherrington-Kirkpatrick model). Extending the theory below its upper critical
dimension $d_\mathrm{u}=6$ is problematic, as explained in
Section~\ref{sec:MODEL}. Hence, it is not simple to guess which of the 
properties of the mean-field solution will remain valid in $d=3$.

Maybe the most prominent feature of the mean-field solution is the resilience
of the spin-glass transition to an external magnetic field (at least, this is
the property that justifies writing this separate paragraph).  The RSB theory
expects a divergence of the correlation length at $T_\mathrm{c}(H)$, the locus
of the de Almeida-Thouless line, in the form of a power law:
\begin{equation}\label{eq:RSB-for-xi}
\xi(T,H)\propto \frac{1}{|T-T_\mathrm{c}(H)|^\nu}\,,
\end{equation}
Consequently, the relaxation time should diverge at $T_\mathrm{c}(H)$. It may
do so in the form of critical slowing down:
\begin{equation}\label{eq:critical-dynamics}
\tau\propto \xi^z\,,
\end{equation}
which defines the dynamic critical exponent, $z$. This is consistent
with the mean-field analysis of the dynamics \cite{sompolinsky:82}.
However, it is possible that the relevant fixed point be at $T=0$,
yet, at variance with the droplet theory, at a finite magnetic field
$H_\mathrm{c}$ (see Figure 1 in \cite{parisi:12}). This is precisely
the situation encountered in the random-field Ising model (see,
e.g., \cite{nattermann:97}), which suggests that an activated
dynamics, as in Eq.~\eqref{eq:droplets-dynamics}, might apply instead 
of~\eqref{eq:critical-dynamics}.
 
At any rate, a defining feature of the RSB picture is the
non-triviality for the probability distribution of the spin overlap $q$,
Eq.~\eqref{eq:q}.  If, for $T<T_\mathrm{c}(H)$, one takes first the limit of
large times $t_{\mathrm{w}}\to \infty$ and only afterwards lets $L\to\infty$,
all values of the overlap in an interval $q_\mathrm{min}\leq q \leq
q_\mathrm{max}$ can be found. Now, we shall be studying the problem with the
reverse order of limits (i.e., first $L\to\infty$, and only afterwards
$t_\mathrm{w}\to\infty$).  Since our simulations will start from a disordered
state, with $q=0$, for temperatures below the dAT line one expects
\begin{equation}
\label{eq:qbis}
 \lim _{t_\text{w}\to \infty} q(t_\text{w})=q_\mathrm{min} \,.
 \end{equation}
From Eqs. (\ref{eq:qbis}) and (\ref{eq:W_term}) one can conclude:
\begin{equation}
\label{eq:diff}
\lim _{t_\text{w}\to \infty} \bigl( W(t_\text{w})-q(t_\text{w})\bigr)=\overline{\langle q
  \rangle}- q_\mathrm{min} \,.
\end{equation}
In the droplet model, the rhs of the previous equation is just zero, while it
is non-zero in a RSB spin-glass phase.

\subsection{A new scenario: supercooled liquids}\label{sec:MCT-and-beyond} 

Probably, the most successful theory for supercooled liquid dynamics is the
Mode-Coupling theory (MCT) \cite{goetze:92,goetze:09}. This is a very rich
theory, with many predictions. We shall content ourselves by recalling the
results most directly relevant to our discussion.

It is now well understood that MCT is a Landau-like or classical
theory \cite{andreanov:09,franz:12a,franz:12b}. As such, it is exact above
eight spatial dimensions. In our $d=3$ world, MCT is adequate only at high
temperatures. Upon lowering the temperature, the correlation length grows to
the point of spoiling the classical critical behavior.  A Ginzburg criterion
can be derived to assess the validity region of MCT \cite{franz:12a,franz:12b}.
A posteriori, the Ginzburg criterion explains the remarkable success of MCT in
the interpretation of dynamical experimental or numerical data for supercooled
liquids.

The MCT predicts a critical divergence of the autocorrelation time at a
temperature $T_\mathrm{d}$
\begin{equation}\label{eq:MCT-1}
\tau\propto \frac{1}{(T-T_\mathrm{d})^\gamma}\,.
\end{equation}
The correlation length diverges as well at $T_\mathrm{d}$ as
\begin{equation}\label{eq:MCT-2}
\xi\propto\frac{1}{(T-T_\mathrm{d})^{1/4}}\,.
\end{equation}
So $\tau\propto \xi^{4\gamma}$ as in Eq.~\eqref{eq:critical-dynamics}. The
critical exponent $\gamma$ is given in terms of two other critical exponents
$a$ and $b$, whose precise definition is not relevant to us now (for instance,
see \cite{franz:12a}):
\begin{equation}\label{eq:MCT-3}
\gamma=\frac{1}{2a}+\frac{1}{2b}\,.
\end{equation}
The two exponents $a$ and $b$ are related to a crucial exponent $\lambda$
through the equation ($\varGamma(x)$ is Euler's $\varGamma$ function~\cite{abramowitz:72}):
\begin{equation}\label{eq:MCT-4}
\lambda=\frac{\varGamma^2(1+b)}{\varGamma(1+2b)}=\frac{\varGamma^2(1-a)}{\varGamma(1-2a)}\,.
\end{equation}
Although $\lambda$ is often treated as an adjustable parameter, we
remark that it is actually a \emph{static} renormalized coupling
constant \cite{caltagirone:12}. As such, it may be systematically computed (for
instance, in an hypernetted-chain approximation \cite{franz:12a,franz:12b} or
in a numerical simulation). A typical value for supercooled liquids is
$\lambda\sim 0.7$ \cite{goetze:09}. We remark that Eq.~\eqref{eq:MCT-4} can be
solved for $a$ and $b$ only if $\lambda\leq 1$. Hence, if one finds
$\lambda>1$ (either in a computation or in a experiment) the problem will have
likely entered a strongly non-perturbative regime.  Overall, MCT is an
adequate theory for the description of the $\beta$-relaxation in liquids.
Roughly speaking this regime covers the temperature range corresponding to $10^2<\tau<10^4$, as
measured in Monte Carlo steps (see, e.g., \cite{kob:94} whose
numerical findings are consistent with $\lambda=0.792$).

However, as we said above, in spite of its success MCT is a Landau
theory. Indeed, at $T_\mathrm{d}$ neither $\tau$ nor $\xi$ diverge.  Instead,
activated processes enter the stage, playing the role of a non-perturbative
phenomenon that erases the mode-coupling transition \cite{lubchenko:07}. The
dynamics is expected to become of activated type, as in
Eq.~\eqref{eq:droplets-dynamics}. The behavior of $\xi$ upon lowering the
temperature is still unclear. Some expect that $\xi$ will diverge at the
Kautzman temperature \cite{mezard:99,mezard:99b,coluzzi:00,parisi:10}, but the
issue is still under active investigation \cite{singh:13}.

%%%%%%%%PROTOCOLS
\section{Dynamic protocols and the identification of the relevant time scale}\label{sec:PROTOCOLS_AND_TAUS}

\subsection{The dynamic protocols}\label{sec:PROTOCOLS}

We have performed several independent sets of simulations, both at a fixed
temperature (\emph{direct quench}) and with temperature changes (\emph{annealing}) . Our
aim was to identify temperature-dependent, intrinsic properties (by
intrinsic we mean independent of the dynamic protocol that we followed).

In all cases we have simulated four replicas for each sample.  We have 
consider external fields $H=0.1$, $0.2$ and $0.3$. 
The linear size of the system is always $L=80$.
We store the full configuration of the system for 
all times of the form $t_\text{w}=[2^{i/4}]$, where $[\cdots]$
denotes the integer part. From these stored configuration we can compute
any physical quantity offline.

In the simulations at a {\em fixed temperature} we ran 462 samples for each
external field at $T=0.7$ and we also simulated 32 samples at 
higher temperatures for $H=0.2$ (see Table~\ref{tab:quench}).
The length of these simulations is $10^{10}$ MCS.

\begin{table}[!ht]
\begin{ruledtabular}
    \begin{tabular}{ccccc}
      $L$ & $T$ & $H$ &MCS & $N$  \\
      \hline
      $80$ & $0.7$ &$0.1$ & $10^{10}$ & $462$ \\
      \hline                     
      $80 $ & $0.7$ & $0.2$ & $10^{10} $ & $462$ \\
      $80 $ & $0.9$ & $0.2$ & $10^{10} $ & $32$ \\
      $80 $ & $1.0$ & $0.2$ & $10^{10} $ & $32$ \\
      $80 $ & $1.1$ & $0.2$ & $10^{10} $ & $32$ \\
      \hline
      $80 $ & $0.7$ & $0.3$ & $10^{10} $ & $462$ \\
    \end{tabular}
\end{ruledtabular}
  \caption{Details of the simulations at fixed temperature. MCS means total
    Monte Carlo sweeps and $N$ is the number of samples
    simulated. \label{tab:quench}}
\end{table}

\begin{table}[!ht]
\begin{ruledtabular}
    \begin{tabular}{ccccccc}
      $L$ & $[T_\mathrm{init}, T_\mathrm{end}]$ & $H$ & $t_\text{base}$ & $\Delta T$ & MCS & $N$  \\
      \hline
      $80$ & $[2.0,0.4]$ &$0.1$ & $10^0$--$10^5$ & 0.1 & $1.3\times10^{10}$ & $995$ \\
      \hline
      $80 $ & $[2.0,0.4]$ & $0.2$ & $10^0$--$10^5$ & 0.1 & $ 1.3\times10^{10}$ & $999$ \\
      $80 $ & $[1.20,0.85]$ & $0.2$ & $512\times10^5 $ & 0.05 & $ 1.3\times10^{10}$ & $1076$ \\
      \hline
      $80 $ & $[2.0,0.4]$ & $0.3$ & $10^0$--$10^5$ & 0.1& $1.3\times10^{10}$ & $1000$ \\
    \end{tabular}
\end{ruledtabular}
  \caption{Details of the simulations with the annealing
    algorithm. The same notation as in Table (\ref{tab:quench}).
   Now $T_\text{ini}$ and $T_\text{end}$ mark the initial and final
    temperatures of the annealing procedure. We decrease the temperature
in $\Delta T$ increments and we run for a time of $t_\text{base}\times
2^{\left(\frac{T_\mathrm{init}-T}{\Delta T}\right)}$ MCS 
at each temperature $T$. For $H=0.2$ we have two sets of 
simulations, the \emph{cold} and the \emph{hot}
annealings. For $H=0.1,0.3$ we only have the cold annealings.
\label{tab:annealing}}
\end{table}

The second set of simulations was performed with an {\em annealing algorithm}.
We started the simulation at a high temperature $T_\text{init}=T_0$.  After $t_\text{base}$
MCS, we change the temperature to a new one $\Delta T$
cooler, i.e., $T_1=T_0-\Delta T$, and we take $2t_\text{base}$ steps.
We iterate this procedure, decreasing the temperature by a fixed step
$\Delta T$ and increasing the number of steps at fixed temperature 
in a geometric progression until we reach our lowest temperature $T_\text{end}$.
That is, for a given temperature $T_k$ the total elapsed time is
in the range $t_k \leq t_\text{tot} \leq t_{k+1}$, where
\begin{align}\label{eq:times}
t_k &= (2^k-1) t_\text{base}.
\end{align}

We performed in every case the annealing from $T_0=T_\text{init}=2.0$ till $T_\text{end}=0.4$ 
with $t_\text{base}=10^i$, $i=0,\ldots,5$ and $\Delta T=0.1$.  We simulated 1000 samples for each
external field, performing a total of $1.3\times10^{5}\times t_\text{base}$
MCS in each sample and replica. See Table \ref{tab:annealing} for more details.
Throughout the paper we shall refer to these runs as the \emph{cold annealings}.
If we do not say otherwise, a mention in the paper to a cold annealing
will always refer to the slowest one, with $t_\text{base}=10^5$.

Finally we have run a yet slower annealing constrained to the high-temperature
region in order to obtain equilibrium results.  In this case, only for $H=0.2$,
we go from $T_\text{init} = 1.2$ down to $T_\text{end}=0.85$, 
with $t_\text{base}=512\times10^5$ and $\Delta T=0.05$. We shall refer 
to these simulations as the \emph{hot annealing}.
\begin{figure}[t]
\begin{center}
  \includegraphics[height=\columnwidth,angle=270]{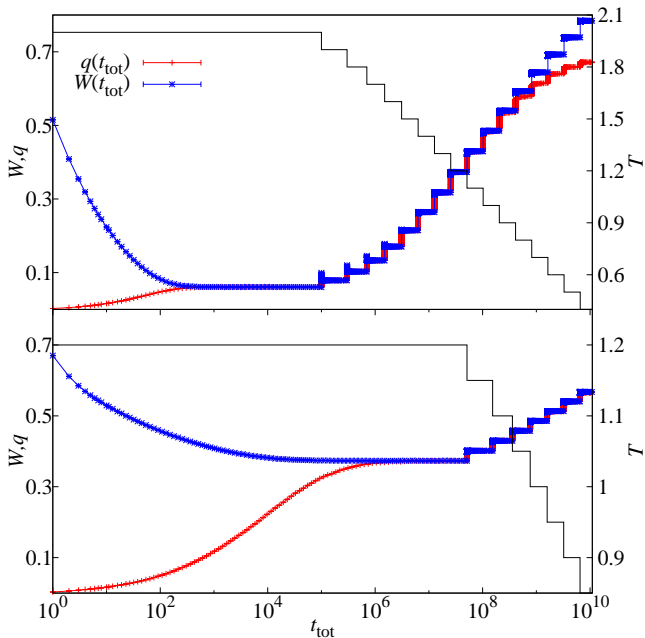}
\end{center}
  \caption{(Color online) Overview of our two annealing simulations for $H=0.2$. We show 
  the evolution of $q(t_\text{tot})$ [the red (lower) curve]
 and $W(t_\text{tot})$ [the blue (uppe) curve] during the whole 
simulation. The continuous black line indicates the temperature throughout 
the simulation (see right-hand vertical axis for the scale). The top panel
corresponds to the \emph{cold} annealing and the bottom panel to 
the \emph{hot} annealing (see Table~\ref{tab:annealing}).
Notice how, in the former, the system is not able to reach equilibrium for
the lowest temperatures (as signaled by different values of $W$ and $q$ at the end of
each temperature step).
   \label{fig:annealing}
}
\end{figure}
\begin{figure}[t]
\begin{center}
  \includegraphics[height=\columnwidth,angle=270]{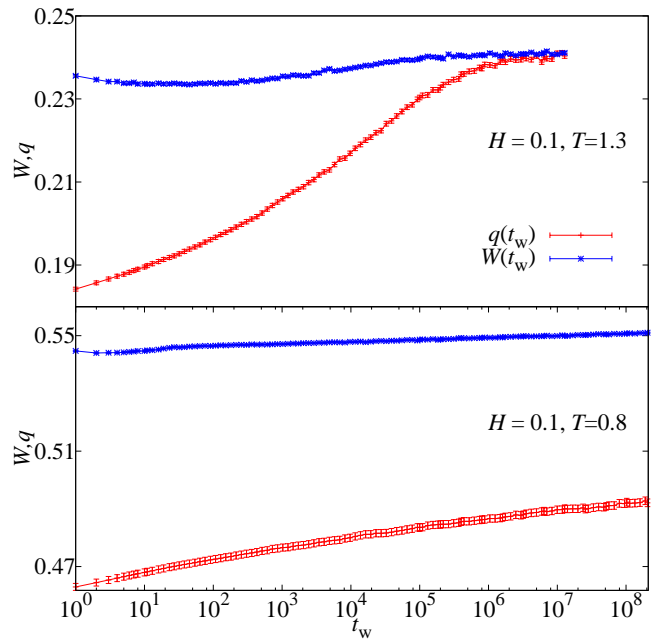}
\end{center}
  \caption{(Color online) 
  Detail of our annealing simulation for $H=0.1$ at two different temperatures, $T=1.3$
and $T=0.8$. Notice how in the second case, despite the longer simulation time, 
the system does not reach equilibrium.
    \label{fig:W-q}
}
\end{figure}

We show  in Figure~\ref{fig:annealing} an overview of our two 
sets of annealing runs for $H=0.2$. We represent both 
$W(t_\text{tot})$ and $q(t_\text{tot})$, which, for 
large $t_\text{w}$, should converge to the same value (in the paramagnetic phase). 
As we can see, in the cold annealing this condition 
is not satisfied for several of the lower temperatures, signaling
that the system has fallen out of equilibrium. The hot annealing
reaches the equilibrium regime for the whole temperature range.

A more detailed picture of these two regimes can be seen
in Figure~\ref{fig:W-q}, which represents $W(t_\text{w})$
and $q(t_\text{w})$ for two temperatures and $H=0.1$.
As we can see, for the higher temperature both observables
converge to the same long-$t_\text{w}$ limit. For the lower
temperature, however, the two quantities are far apart 
during the whole simulation and seem to have a different asymptote.
This indicates that either the equilibration time is much larger 
than our simulation or we are in a spin-glass phase.
Deciding between these two possibilities is the main goal of 
this paper.

\subsection{Identification of intrinsic time scales}\label{sec:TAUS}

\begin{figure}[t]
  \includegraphics[height=\columnwidth,angle=270]{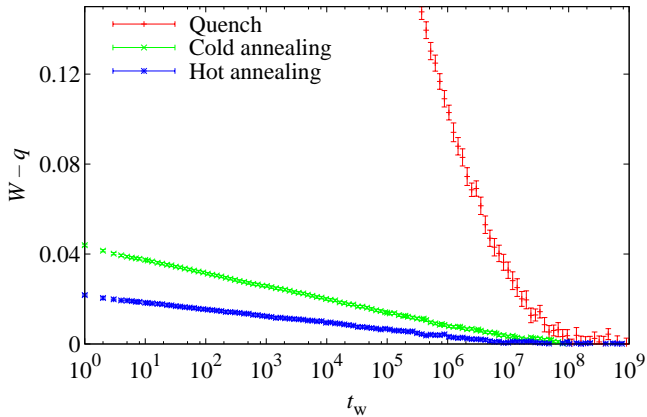}
  \caption{(Color online)
Our three different simulations at $T=1.0$ for $H=0.2$. We show the difference
$W-q$ as a function of the waiting time $t_\text{w}$, whose decay to zero
determines the time scale of the simulations. Despite having
very different initial conditions, the three simulation protocols need roughly 
the same time to reach equilibrium.
We shall quantify this assertion in this Section~\ref{sec:TAUS}.
Notice how, for the two annealing protocols, the behavior
of $W-q$ is linear in $\log(t_\text{w})$ for a long time range.
    \label{fig:W-q-T1}
}
\end{figure}

Our strategy will rely on the study of the difference between $W(t_\text{w})$ and
$q(t_\text{w})$ as a function of time, which should decay to zero
in the high-temperature phase. Some inspiration comes from the classic paper
by Ogielski \cite{ogielski:85}, although we shall use other approaches
as well.

We show in Figure~\ref{fig:W-q-T1} the behavior of
$W(t_\text{w})-q(t_\text{w})$ for $T=1.0$ and $H=0.2$ using the three
simulations protocols described in the previous section (direct quench and
cold and hot annealing). Not surprisingly, the starting value of $W-q$
differs greatly from one protocol to the other. It is very large (well out of
the graph's scale) for the direct quench, which starts with a random
configuration (representing a very high temperature) and it is smallest for
the hot annealing, which has a small temperature step.

Nevertheless, all protocols seem to need roughly the same number
of MCS to reach equilibrium, as evinced by the merging of the curves 
at $t_\text{w}\sim 10^8$. This observation gives us some 
hope of determining an intrinsic time scale $\tau$, depending 
only on the system's temperature and not on its
history.

In principle, the robust way to compute $\tau$ would be to perform
a calculation analogous to Eq.~\eqref{eq:xi_integrals}, replacing
$C$ by $W-q$ and $r$ by $t_\text{w}$. Unfortunately, in the interesting 
temperature range the maximum of the integrand $t_\text{w}^k [W-q](t_\text{w})$
is always in a region with a dismal signal-to-noise ratio (cf. Figure~1 
in \cite{janus:09b}). Therefore, we have to resort to more
phenomenological determinations.

A traditional way to identify this time scale,  which 
was found adequate in the absence of a field \cite{ogielski:85},
is fitting the difference $W(t_\text{w})-q(t_\text{w})$ to
a stretched exponential decay:
\begin{equation}
W(t_\text{w})-q(t_\text{w})=\frac{\mathcal A}{t_\text{w}^x} \exp\left[-\left(\frac{t_\text{w}}{\tau'}\right)^\beta\right] \,.
\label{eq:dif_hT}
\end{equation}
From this fit one gets a characteristic time $\tau'$, which we can use
as our time scale.

Computing a fit to Eq.~\eqref{eq:dif_hT} is difficult due to the extreme
correlation of our data, which prevents us from inverting its full covariance
matrix (necessary to define the $\chi^2$ goodness-of-fit indicator). Therefore,
we consider only the diagonal part of the matrix in order to minimize $\chi^2$
and take correlations into account by repeating this procedure for each
jackknife block in order to estimate the errors in the parameters. This is, of
course, only an empirical procedure, but one that has been shown to work well
under these circumstances (see, e.g., \cite{janus:09b}, especially sections 2.4
and 3.2). 

The results of these fits are gathered in Table~\ref{tab:tau}. We do not 
report the value of the (diagonal) $\chi^2$/d.o.f. because it is, in all 
cases, $\chi^2/\text{d.o.f.}<1$ (as we have said this indicator does not
give the full picture in the presence of strong data correlations). 
For each value of the magnetic field we have fitted up to the point
where the system falls out of equilibrium (as indicated by a $\tau'$ longer
than the simulation time).
\begin{table*}[t]
\begin{ruledtabular}
\begin{tabular}{ccclccll}
\multirow{2}{*}{$H$} & \multirow{2}{*}{$T$} &\multirow{2}{*}{Annealing} & \multicolumn{2}{c}{Stretched exponential, Eq. \eqref{eq:dif_hT}} &  & \multicolumn{2}{c}{Linear fit, Eq. \eqref{eq:logtau}}\\
\cline{4-5} \cline{7-8}
& & &\multicolumn{1}{c}{$\tau'/10^7$} & \multicolumn{1}{c}{$\beta$} & & \multicolumn{1}{c}{$\log(\tau'')$} & \multicolumn{1}{c}{$\tau''/10^9$}\\ 
\hline
\multirow{3}{*}{0.3} & 1.0 & Cold & 0.015(7) & 0.27(4) & & 14.69(12) & 0.0024(3)\\
                     & 0.9 & Cold & 0.09(4)   & 0.23(2) & & 17.20(15) & 0.029(4)\\
                     & 0.8 & Cold & 3.2(11) & 0.21(2) & & 20.71(19) & 0.99(19)\\
\hline
\multirow{6}{*}{0.2} & \multirow{2}{*}{1.1} & Cold       & 0.022(7)   & 0.30(4) & & 15.12(14)  & 0.0037(5) \\
                   &  & Hot       & 0.027(10)   & 0.37(6) &  & 14.60(27)  & 0.0022(6)  \\
\cline{2-8}
& \multirow{2}{*}{1.0} & Cold       & 0.19(7) & 0.26(3)   & & 17.37(19) & 0.035(7) \\
                  &   & Hot       & 0.14(6)  & 0.31(7) &   & 16.9(3) & 0.022(7)\\
\cline{2-8}
& \multirow{2}{*}{0.9} & Cold       & 5.1(1.9)& 0.23(3) &     & 20.81(26) & 1.1(3) \\
                  &   & Hot       & 3.1(2.2)  & 0.23(5) &    & 20.2(6) & 0.6(4) \\
\hline
\multirow{3}{*}{0.1} & 1.4 & Cold & 0.0012(3) & 0.40(4) &  & 11.73(12) & 0.000125(16)\\
                     & 1.3 & Cold & 0.006(3)   & 0.32(5) &  & 14.08(14) & 0.00130(18)\\
                     & 1.2 & Cold & 0.060(14) & 0.34(3) &  & 16.44(21) & 0.014(3)\\
\end{tabular}
\end{ruledtabular}
\caption{Computation of characteristic times for several temperatures
and simulation protocols using the stretched exponential~\eqref{eq:dif_hT}
as well as a linear fit in $\log(t_\text{w})$~\eqref{eq:logtau} ($\log$ being
the natural logarithm).
For each value of $H$ we report the last few temperatures before the system
falls out of equilibrium.
All the reported fits have (diagonal) $\chi^2/\text{d.o.f.} < 1$.
\label{tab:tau}}
\end{table*}
\begin{figure}[t]
\begin{center}
  \includegraphics[height=\columnwidth,angle=270]{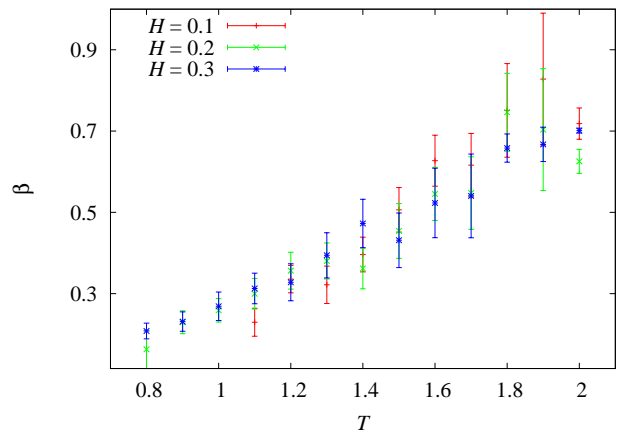}
\end{center}
  \caption{(Color online) Behavior of the stretching exponent $\beta$ (see Eq.\eqref{eq:dif_hT})
as a function of $T$ for our three simulated magnetic fields.
  \label{fig:beta}}
\end{figure}

A possible source of uncertainty
in our determination of $\tau'$ is the dependence of the fit on the value 
of $\beta$. Indeed, for each $T$ we are fitting simultaneously for $x$, $\mathcal A$, 
$\tau'$ and $\beta$ in~\eqref{eq:dif_hT}. However, a small variation in $\beta$ 
can have a large effect on $\tau'$, which may lead us to think that the fit is
unstable and unreliable. Fortunately (see Fig.~\ref{fig:beta}), $\beta$ is 
actually a very smooth monotonic function of $T$, which leads us to believe
that the fitting procedure is sound.

There is a final difficulty with this functional form:
$\tau'$ only has a straight interpretation as a
correlation time (i.e., as an estimator for $\tau$)
if $\beta\approx 1$. However, in the interesting 
temperature range, $0.2\lesssim \beta\lesssim 0.3$. 
This means that the actual value of $\tau'$ cannot
be interpreted directly as an estimator for $\tau$,
 but still we expect its divergence 
at the dynamical transition point  to be intrinsic, 
as discussed in Section~\ref{sec:equilibrium}.

Notice, finally, that the values of $\tau'$ and $\beta$ computed at the same 
temperature for the hot and cold annealing protocols 
for $H=0.2$ are compatible. This is a very good indication
that these parameters have some intrinsic meaning.
\begin{figure}[t]
\begin{center}
  \includegraphics[height=\columnwidth,angle=270]{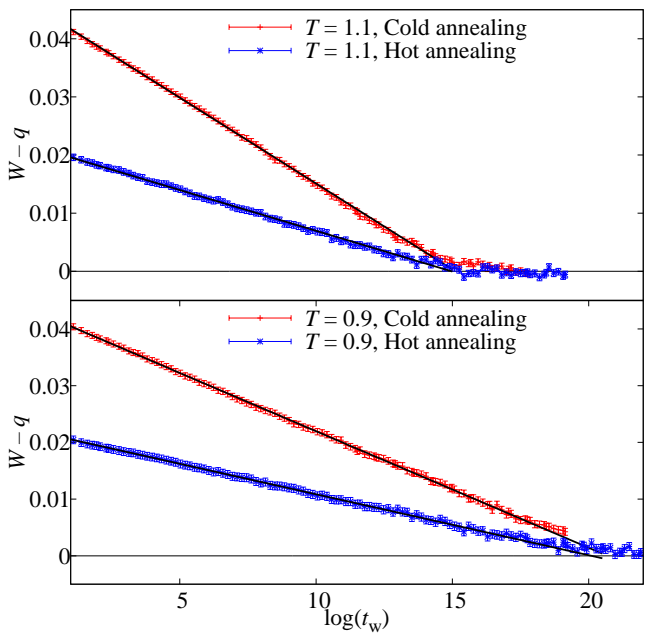}
\end{center}
  \caption{(Color online)
  Detail of our two annealing simulations for $H=0.2$ at two different temperatures, $T=1.1$
and $T=0.9$.  We represent the difference $W-q$ as a function of
$\log(t_\text{w})$ (we have used $\log$ as the natural logarithm throughout the
paper). As 
pointed out in Figure.~\ref{fig:W-q-T1}, in this representation $W-q$ is linear 
for a very long range. More interestingly, if we compute linear fits to~\eqref{eq:logtau}
(continuous black lines), the intercepts with the horizontal axis are independent 
of the simulation protocol. 
    \label{fig:W-q-log}
}
\end{figure}

Another approach to the estimation of $\tau$, completely
phenomenological, comes to mind from a visual
inspection of Figure~\ref{fig:W-q-T1}. Indeed, 
we can see that for both annealing protocols, 
the difference $W-q$ is linear in $\log(t_\text{w})$
for a very wide temperature range. This behavior
is more clearly shown in Figure~\ref{fig:W-q-log}, which
represents this quantity for two different temperatures.

Moreover, if we represent this linear behavior as 
\begin{equation}\label{eq:logtau}
W(t_\text{w})-q(t_\text{w}) \simeq \mathcal B\left[1 - \frac{\log(t_\text{w})}{\log(\tau'')}\right],
\end{equation}
we can see from Figure~\ref{fig:W-q-log} that the value of $\tau''$ 
does not depend on the annealing rate and is therefore dependent 
only on the temperature. Furthermore, since there
is no stretching exponent, we can take $\tau''$ directly as 
an estimate of the actual intrinsic time scale 
of the system: $\tau''\approx \tau$.

Of course, Eq.~\eqref{eq:logtau} is only empirical and cannot 
be correct for very long times (it would predict an unphysical 
negative value of $W-q$ for $t_\text{w}>\tau''$), but its simplicity
and robustness compensate for this problem. We give
the values of $\tau''$ for several temperatures in Table~\ref{tab:tau} (in accordance 
with the previous discussion on the meaning of $\beta$, notice that $\tau''$ is more
than an order of magnitude larger than $\tau'$).
In the following, we shall use both $\tau'$ and $\tau''$ to study the possible critical
behavior of the system.
 
%%%%%%EQUILIBRIUM
\section{The equilibrium regime}~\label{sec:equilibrium}
\begin{figure}[t]
\begin{center}
  \includegraphics[height=\columnwidth,angle=270]{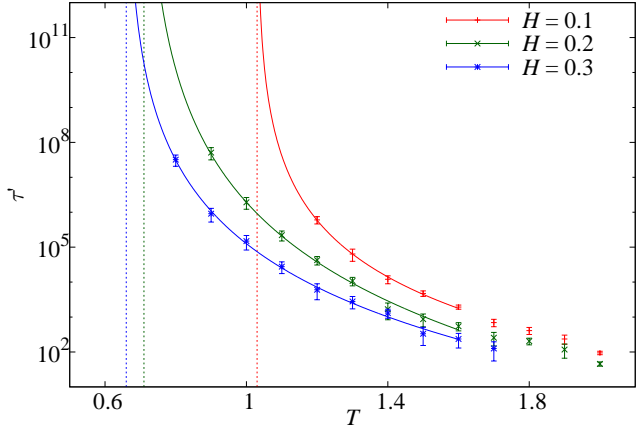}
\end{center}
  \caption{(Color online) Behavior of the characteristic time $\tau'$ of the stretched exponential~\eqref{eq:dif_hT}
    as a function of the
    temperature for the three magnetic fields simulated. 
     We also plot our fits to a power-law divergence of $\tau'$ for
     a finite temperature, Eq.~\eqref{eq:div_tau}.    
  \label{fig:T_tau}}
\end{figure}
In this section we consider the temperature dependence of the
characteristic time scales $\tau$ identified in Sect.~\ref{sec:TAUS}.
Following Ogielski~\cite{ogielski:85} as well as experimental studies
(e.g., \cite{jonsson:00}), we shall fit our equilibrium data to
power-law divergence:
\begin{equation}
\tau=\frac{\tau_0}{\bigl[T-T_\text{c}(H)\bigr]^{\nu z}}\,,
\label{eq:div_tau} 
\end{equation}
where $\tau_0$ is a microscopical time.
We use the traditional notation, where
$T_\text{c}(H)$ is a critical temperature, 
$\nu$,  is the correlation length critical exponent and $z$
is the dynamical critical exponent. However, as we shall see
in Section~\ref{sec:xi}, this divergence may or may not correspond
to an actual phase transition at $T_\text{c}(H)$.

In Figure~\ref{fig:T_tau} we show the relaxation time $\tau'$ computed with
the stretched exponential~\eqref{eq:dif_hT} in Table~\ref{tab:tau} as a
function of temperature for the three simulated magnetic fields. We also show
fits to the power-law divergence at finite $T_{\mathrm{c}}^{\mathrm{high}}(H)$
(the superscript ``high'' refers to the fact that we are using
high-temperature data, cf. Section~\ref{sec:NON-EQUILIBRIUM}). We have
obtained the following values:
\begin{itemize}
\item $H=0.1$: $T_\text{c}^\text{high}=1.03(7)$ and $z \nu=4.8(1.1)$. Using only  $1.2\le T \le 1.7$ [$\chi^2/\mathrm{d.o.f.}=1.47/3$].
\item $H=0.2$: $T_\text{c}^\text{high}=0.71(6)$ and $z \nu=7.5(1.1)$. Using only $0.9 \le T \le 1.6$ [$\chi^2/\mathrm{d.o.f.}=3.36/5$].
\item $H=0.3$: $T_\text{c}^\text{high}=0.66(5)$ and $z \nu=6.2(9)$. Using only $0.8 \le T \le 1.5$ [$\chi^2/\mathrm{d.o.f.}=1.74/5$].
\end{itemize}
In all the fits we go down to the lowest temperature where we can measure $\tau'$ reliably. 
The choice of the fitting range is not critical, several temperatures can be added or eliminated 
without altering the results significantly (especially in the high-temperature end).
For $H=0.2$, we use the $\tau'$ for the cold annealing, since they have smaller error bars
than those for the hot annealing. We have checked that the $\tau'$ at $T=0.85$ extrapolated with 
the above fits for the cold annealing is compatible with the corresponding correlation time  
measured in the hot annealing.

However, as we discussed in the previous section, $\tau'$ does not have a straightforward interpretation
as a relaxation time, since $\beta \neq 1$. Therefore, the above values of $T_{\mathrm{c}}^{\mathrm{high}}$ might be an artifact 
of our way of estimating $\tau$. In order to dispel this possibility, we have 
recomputed the fits to~\eqref{eq:div_tau}, this time using the relaxation time $\tau''$ 
computed with the linear fit in $\log t_\text{w}$~\eqref{eq:logtau}. Now the fit parameters are:
\begin{itemize}
\item $H=0.1$: $T_\text{c}^\text{high}=0.98(3)$ and $z \nu=7.2(5)$. Using only $1.2\le T \le 1.7$ [$\chi^2/\mathrm{d.o.f.}=4.21/3$].
\item $H=0.2$: $T_\text{c}^\text{high}=0.670(21)$ and $z \nu=9.2(4)$. Using only $0.9 \le T \le 1.6$ [$\chi^2/\mathrm{d.o.f.}=1.79/5$].
\item $H=0.3$: $T_\text{c}^\text{high}=0.614(17)$ and $z \nu=8.4(4)$. Using only $0.8 \le T \le 1.3$ [$\chi^2/\mathrm{d.o.f.}=2.61/3$].
\end{itemize}

We can see that we obtain good values of the goodness-of-fit estimator $\chi^2$
for all fits.  The values 
of $T_\text{c}(H)$ are consistent for both sets of fits, while the values of $z\nu$ are a little higher
for the fit with $\tau''$ ($2.0$ standard deviations for $H=0.3$, $1.5$ standard deviations
for $H=0.2$ and $2.2$ standard deviations for $H=0.1$).

The consistency between these two sets of fits makes us confident that the observed divergence 
in the relaxation times is an intrinsic phenomenon and not an artifact of 
our simulation protocol or of our ansatz for the behavior of $W-q$.

\subsection{The relaxation time in the supercooled liquids approach}\label{sec:tau-mct}
As discussed in Section~\ref{sec:MCT-and-beyond}, the MCT predicts a
divergence of the autocorrelation time at a temperature $T_\text{d}$, as in
Eq.~\eqref{eq:MCT-1}.  In principle, Eq.~\eqref{eq:MCT-1} is exactly the same
as Eq.~\eqref{eq:div_tau}, which we have just used to characterize the growth
of the relaxation times.  The crucial difference is that we have used our
lowest thermalized temperatures and  have assumed that the growth of $\tau$ was
related to an actual critical divergence (as evinced by our notation of $z\nu$
for the exponent). On the other hand, in the supercooled liquids literature, Eq.~\eqref{eq:MCT-1} is
used in a higher temperature range corresponding to $10^2<\tau<10^4$ 
(notice, for instance, that our values of $z\nu$ are very high compared 
to the values of $\gamma$ that can be found in the MCT literature). For lower
temperatures, the behavior of $\tau$ deviates from~\eqref{eq:MCT-1}, because of
the emergence of activated processes. 

Therefore, if we wanted to follow a supercooled liquids approach, we should first fit $\tau'$
to~\eqref{eq:MCT-1} in the high temperature range and then move on to an exponential 
growth:
\begin{align}
\tau' &= \frac{\mathcal C}{(T-T_\text{d})^\gamma}, &  \tau' \lesssim 10^4,\label{eq:MCT-tau-1}\\
\tau' &= \exp\bigl[ \mathcal{D}/ (T-T_\text{c})^{\gamma'}\bigr], & \tau' \gtrsim 10^4. \label{eq:MCT-tau-2}
\end{align}
Unfortunately, our simulations are not suited to the determination of small $\tau$, so
the fit to~\eqref{eq:MCT-tau-1} will probably be plagued by strong systematic effects.

We consider only $H=0.2$ (for $H=0.1$ we have too narrow a temperature range and for $H=0.3$ 
our fits for $\tau'$ are rather unstable for $T>1.7$). We have fitted $\tau'$ 
to~\eqref{eq:MCT-tau-1}  in the range $1.3\leq T\leq1.9$ (which corresponds to $10^2<\tau'<10^4$).
The result is $T_\text{d}=1.22(6)$ with $\gamma=2.1(7)$ ($\chi^2/\text{d.o.f.}=0.94/4$). 
The next step would be to take $\tau'$ in the range $T\leq1.3$ (which we have
previously fitted successfully to a critical divergence with $z \nu\approx 8$)
and attempt a fit to~\eqref{eq:MCT-tau-2} instead. Unfortunately, fitting
for $T_\text{c}$ and $\gamma'$ simultaneously is simply not possible with our data
(the resulting error in $T_\text{c}$ is greater than $100\%$). In short, we can only 
say that a temperature dependence of $\tau$ according to~\eqref{eq:MCT-tau-1} and~\eqref{eq:MCT-tau-2}
cannot be excluded, but we cannot make this statement more quantitative. Nevertheless, we 
shall return to the possibility of activated dynamics in Sections~\ref{sec:NON-EQUILIBRIUM-LINEAR-LOG}
and~\ref{sec:xi}.

%%%%%%%NON-EQUILIBRIUM
\section{Non-equilibrium regime}\label{sec:NON-EQUILIBRIUM}

As already explained in Section~\ref{sec:TAUS}, our annealing rate eventually
becomes too fast, compared to the growth of $\tau(T)$ upon cooling. At that
point, the simulation falls out of equilibrium. We enter here the reign of
extrapolation, which is always rather risky.

We shall extrapolate our data to long times following two very different
strategies. In Section~\ref{sec:NON-EQUILIBRIUM-POWER-LAW} we extrapolate
using power laws. The outcome will be consistent with the RSB theory. On the
other hand, in Section~\ref{sec:NON-EQUILIBRIUM-LINEAR-LOG} we use  the
linear-log extrapolation (see Sect.~\ref{sec:TAUS}), which assumes from
the outset that no phase transition occurs.

\subsection{Power-law extrapolations to long times}\label{sec:NON-EQUILIBRIUM-POWER-LAW}
%---------------------------------------------
So far, we have been working in the high-temperature phase, where
$W-q$ goes to zero for long times.  We saw that, as we 
lower the temperature, the associated relaxation time grows
very quickly and eventually becomes much larger than our 
simulations. This rapid growth was actually consistent 
with a power-law divergence of $\tau$ at finite $T$.

In this section we shall take a complementary approach. We now
work in the low-temperature regime, where $\tau$ is either
infinite or, at the very least, much larger than our simulation times.
In this regime, rather than assuming that $W-q$ goes to zero for 
long times, we can try to extrapolate for a (possibly) non-zero asymptote.

Following the literature~\cite{parisi:98b,marinari:98e}, we shall first attempt a study in 
the total annealing time $t_\text{tot}$, considering different
annealing rates (Section~\ref{sec:power-ttot}). Then we shall repeat 
the analysis using only $t_\text{w}$ (as in the rest of the paper) in Section~\ref{sec:power-tw}.

\subsubsection{Study in $t_\text{tot}$ for different annealing rates} \label{sec:power-ttot}
In the limit of a very slow annealing,
the simplest ansatz for the low-temperature behavior of $W-q$
is a power-law decay (cf. \cite{ogielski:85}):
\begin{equation}
W(t_\text{tot})-q(t_\text{tot})=a(T,H)+\frac{b}{t_\text{tot}^x}\,.
\label{eq:dif_lT}
\end{equation}
Eventually $b$ and $x$ could also depend on the temperature \cite{ogielski:85}
and on the external magnetic field. Notice that, in contrast with the rest
of the paper, here we are considering the total time $t_\text{tot}$ since
the simulation started~\cite{parisi:98b,marinari:98e}, not just the time $t_\text{w}$ since 
the last temperature change.

Should the system experience a spin-glass transition, we would expect
$a(T,H)>0$ for very low $T$ [recall Eq.~\eqref{eq:diff}]. This asymptote would
decrease as we increase the temperature until eventually, at some temperature
$T_\text{c}^\mathrm{low}(H)$, $a\bigl(T_\text{c}^\mathrm{low}(H),H\bigr)=0$. This
description is consistent with a qualitative look at $W-q$ (recall, for
instance, Figure~\ref{fig:W-q}).

If the RSB picture is correct, we would expect $T_\text{c}^\mathrm{low}(H)$ to
coincide with the divergence of $\tau$ and signal a thermodynamic phase
transition, that is, $T_\text{c}^\mathrm{low}(H)=T_\text{c}^\mathrm{high}(H)=T_\text{c}$.

Equation~\eqref{eq:dif_lT}, with $a> 0$, should hold only deep in
the spin-glass phase. If we approach the transition from below (in
temperature) we would start to see the critical effects of the
(thermodynamical) critical point, and the exponent $x$ would begin to be
controlled by this critical point and not by the ``critical'' spin-glass phase
(Goldstone phase).  So, in the critical region we should expect:
\begin{equation}
W(t_\text{tot})-q(t_\text{tot})=\frac{f}{t_\text{tot}^{x_c}} \,,
\label{eq:crit}
\end{equation}
where in general $x_c$ (driven by the critical point) 
\footnote{$x_c$ can be expressed as: $x_c=(d-2+\eta)/(2 z)$, where $d$ is the
  dimensionality of the system, $z$ is the dynamical critical exponent and
  $\eta$ is the anomalous dimension.}  should be different from $x$ (driven by
the spin-glass phase).

From the previous discussion, and assuming the onset of a phase transition, it
is clear that the $x$ exponent should take a constant value at lower
temperatures (here we are assuming that the phase transition is universal in
the magnetic field), then change as we reach the critical region. 

\begin{figure}[t]
\begin{center}
\includegraphics[height=\columnwidth,angle=270]{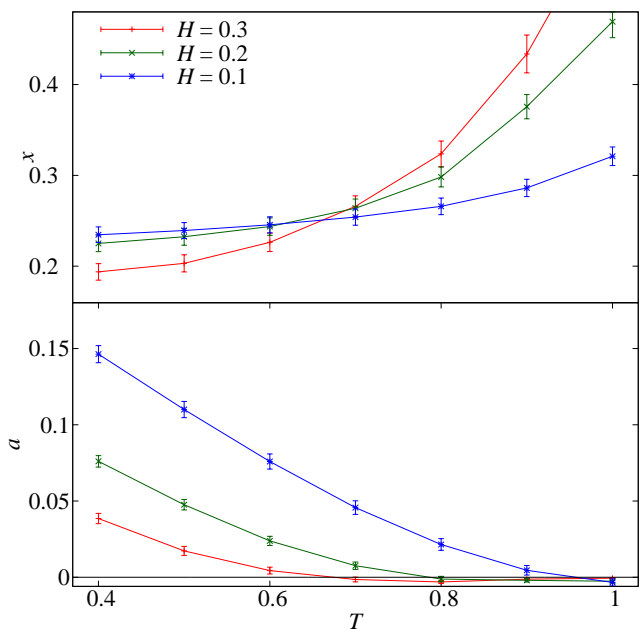}
\end{center}
  \caption{(Color online) \emph{Top:} Exponent $x$ of the extrapolation of the
    difference between $W(t_\text{tot})$ and $q(t_\text{tot})$,
    Eq.~\eqref{eq:dif_lT}, as a function of temperature, for our three
    external magnetic fields. Value computed from three-parameter fits.
    \emph{Bottom:} As above, for the asymptote $a$.   \label{fig:WqT_ttot}}
\end{figure}

Now, in order to study the decay of $W-q$ we obviously need to follow the time
evolution along several orders of magnitude. However, as described in
Section~\ref{sec:PROTOCOLS}, for a fixed temperature $T_k=T_\text{init} - k
\Delta T$ the total time elapsed in our annealing simulations varies in the
range $t_k(t_\text{base}) \leq t_\text{tot} \leq t_{k+1}(t_\text{base}) \sim 2
t_k (t_\text{base})$, which is too narrow in a logarithmic scale. Therefore,
instead of analyzing the data for all the $t_\text{tot}$ in a given annealing
simulation, we shall combine all our cold annealing simulations for different
values of $t_\text{base}$. That is, for each temperature $T_k$ we take the
value of $W-q$ at $t_{k+1}(t_\text{base})$, for $t_\text{base}= 10^i$,
$i=0,\ldots,5$.  Thus, we get for each temperature $T_k$ a series of six values of
$[W-q](t_\text{tot})$ in the range $t_{k+1}(t_\text{base}\!=\!1) \leq t_\text{tot} \leq t_{k+1}(t_\text{base}\!=\!10^5)$,
which we fit to~\eqref{eq:dif_lT} [recall that $t_{k+1}(t_\text{base})=(2^{k+1}-1)t_\text{base}$].

The resulting values of $a(T)$ and $x(T)$ are plotted in
Figure~\ref{fig:WqT_ttot}.  As we can see, the qualitative picture is very much
what we painted above. In particular, we obtain a positive value of the
asymptote $a$ for low temperatures, which goes to zero at a temperature
$T_\text{c}^\text{low}$ not very different from $T_\text{c}^\text{high}$ ---we
can estimate $T_\text{c}^\text{low}(H=0.3)\approx0.65(5),
T_\text{c}^\text{low}(H=0.2)\approx 0.80(5),
T_\text{c}^\text{low}(H=0.1)\approx 0.96(5)$.  In addition, the value of the
exponent $x$ is roughly constant (and independent of $H$) at low temperatures,
while it grows noticeably as we approach $T_\text{c}^\text{low}$.

However, we must caution the reader that the fits we have just discussed are
rather delicate. In particular, even after discarding the two smallest
$t_\text{tot}$ (so we are left with a three-parameter fit to four points), we
find that values of the $\chi^2$ goodness-of-fit estimator are sometimes very
high.  In particular, the fits for $H=0.1,0.3$ are good in the interesting
temperature range (always $\chi^2/\text{d.o.f.}\leq 1.5/1$), but those for
$H=0.2$ have $\chi^2/\text{d.o.f.}$ that can be in excess of $8/1$, clearly
unacceptable.  More worryingly, if we shift the fitting window we find that the
fitted values for $x$ and $a$ decrease noticeably with increasing
$t_\text{tot}$ (the change in $x$ can be as high as $50\%$ just by shifting the
fitting window so that we discard the longest time but include an extra point
in the lower end of the range). Still, for $H=0.1,0.3$ only the fitting window
for the longest $t_\text{tot}$ gives reasonable fits (for $H=0.2$ the situation
is murkier, since the fits are poor in any case).  It will be interesting to
compare these values with the ones we shall obtain in the next section
(Section~\ref{sec:power-tw}), with a study in $t_\text{w}$.

Of course, from the above arguments one could think that, for long enough $t_\text{ttot}$, 
the exponent could decrease so much that the asymptote $a$ would become zero.
In order to check against that possibility, and to obtain a sort of lower bound
for $a$, we have also attempted fits to the following function
\begin{equation}\label{eq:WqT_lT_log}
W(t_\text{tot})-q(t_\text{tot}) = a' + \frac{b'}{\log(t_\text{tot})^{x'}}.
\end{equation}
Using this fitting function we get good fits with values of $x'\approx 3$ 
and positive $a'$ for $T<T_\text{c}^\text{high}$, so the qualitative picture
is the same. The determination of $T_\text{c}^\text{low}$ is a little 
lower,  but still compatible with our $T_\text{c}^\text{high}$.

\begin{figure}[t]
\begin{center}
\includegraphics[height=\columnwidth,angle=270]{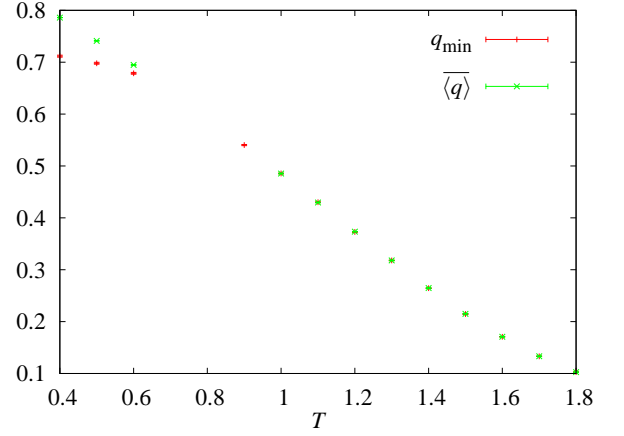}
\end{center}
  \caption{(Color online) Extrapolations to infinite time of $W(t_\text{tot})\to\overline{\langle q\rangle}$ 
and $q(t_\text{tot})\to q_\text{min}$ for $H=0.2$, according to~\eqref{eq:qT}. We use the same exponent $x$ computed
in Figure~\ref{fig:WqT_ttot}.
  \label{fig:qT}}
\end{figure}
The next step in this study is the investigation of the scaling in 
$W$ and $q$ separately. To this end, we are going to consider fits of the form
\begin{equation}\label{eq:qT}
\begin{split}
W(t_\text{tot}) &= \overline{\langle q\rangle} + \frac{b''}{t_\text{tot}^x},\\
q(t_\text{tot}) &= q_\text{min} + \frac{b'''}{t_\text{tot}^x},
\end{split}
\end{equation}
where for both observables we take the same value of $x$ that we computed in
Figure~\ref{fig:WqT_ttot}. The resulting plots of $\overline{\langle
q\rangle}(T)$ and $q_\text{min}(T)$ can be seen in Figure~\ref{fig:qT} for
$H=0.2$. Except for $H=0.2$, we obtain excellent fits both for $q$ and for $W$
($\chi^2 /\text{d.o.f.}< 1$).  For high temperatures ($T\geq0.9$ for $H=0.2$) we do not need
to extrapolate, since we reach equilibrium in our annealing simulations. There
is a small gap with no extrapolations, corresponding to temperatures above
$T_\text{c}$, where~\eqref{eq:dif_lT} does not work but we cannot reach
equilibrium.  The qualitative picture is what one would expect 
in the RSB scenario.

\begin{figure}[t]
\begin{center}
\includegraphics[height=\columnwidth,angle=270]{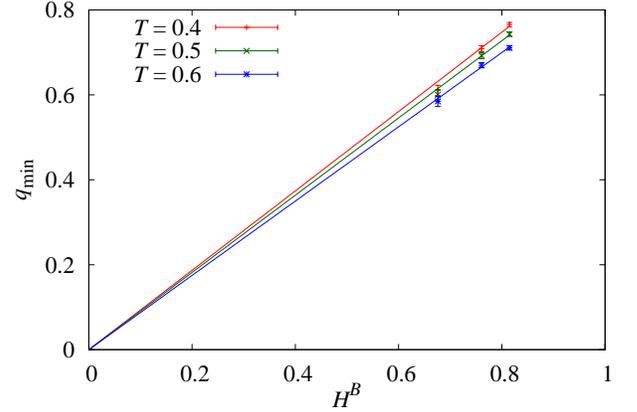}
\end{center}
  \caption{(Color online) Scaling of the minimum overlap with the magnetic field.
   A simple power-law behavior, $q_\text{min} \sim H^B$, works well for 
   all our subcritical temperatures, even though we are far from the small-$q_\text{min}$
   limit.
  \label{fig:qH}}
\end{figure}
Finally, we can consider the scaling of $q_\text{min}$ with the magnetic field at fixed 
temperature. As the magnetic field goes to zero, so must $q_\text{min}$ and we could expect a 
behavior of the kind $q_\text{min}(H) \sim H^B$. Indeed, a rough dimensional analysis \cite{marinari:98e}
tells us that $B=\theta(0)/[D-\theta(0)/2]$, where the replicon exponent in $D=3$ is 
$\theta(0)=0.39(5)$ \cite{janus:10b}.  Therefore, we expect  a value of $B= 0.14(2)$. 
We have computed fits to
\begin{equation}\label{eq:qH}
q(H) = \mathcal C H^B
\end{equation}
for $T=0.4,0.5$ and $0.6$ (the only temperatures that are below $T_\text{c}$ for our three 
magnetic fields). The results are $B(T=0.4) = 0.20(2)$, $B(T=0.5)=0.19(2)$, $B(T=0.6)=0.17(2)$,
very close to our expected $B=0.14(2)$ (notice that we are considering rather high
magnetic fields, as evinced by the high values of $q_\text{min}$ that we are seeing).
In all cases we obtain excellent values of the $\chi^2$ estimator. 
We have plotted $q_\text{min}^B$ in Figure~\ref{fig:qH}, using for all fields an intermediate 
value of $B=0.18$.

\subsubsection{Study in $t_\text{w}$} \label{sec:power-tw}
As discussed above, the total time $t_\text{tot}$ since the simulation started
is the more physical variable to conduct the low-temperature study. However, we have 
seen in Sections~\ref{sec:PROTOCOLS} and~\ref{sec:equilibrium} that
we can also study the relaxation of the system in $t_\text{w}$ in a consistent 
way. Therefore, it is interesting to repeat the study of Section~\ref{sec:power-ttot}
taking only our slowest cold annealing (with $t_\text{base}=10^5$) and studying for each 
temperature the relaxation in $t_\text{w}$.

\begin{figure}[t]
\begin{center}
\includegraphics[height=\columnwidth,angle=270]{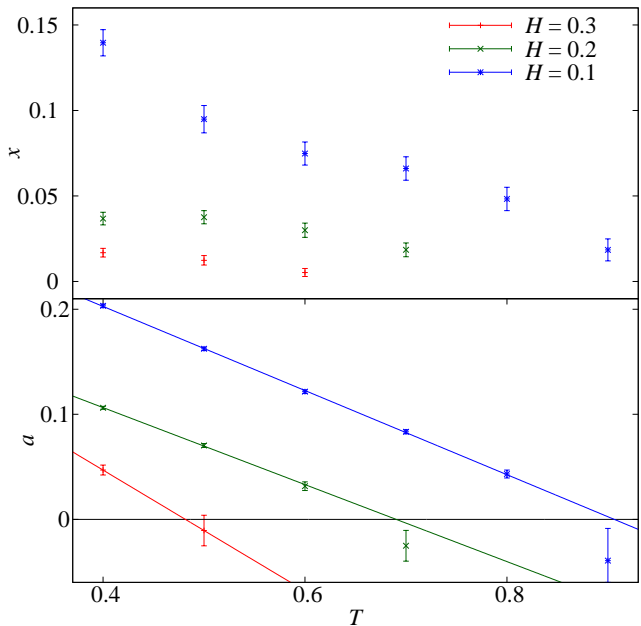}
\end{center}
  \caption{(Color online) \emph{Top:} Exponent $x$ of the extrapolation of the
    difference between $W(t_\text{w})$ and $q(t_\text{w})$,
    Eq.~\eqref{eq:dif_lT_tw}, as a function of temperature, for our three
    external magnetic fields. Values computed in a three-parameter fit.
    \emph{Bottom:} As above, for the asymptote $a$. Lines are linear fits to
    the points where $a(T,H)>0$ (for $H=0.3$ we included as well $T=0.5$ in
    the fit).
  \label{fig:WqT_tw}}
\end{figure}
\begin{figure}[t]
\begin{center}
\includegraphics[height=\columnwidth,angle=270]{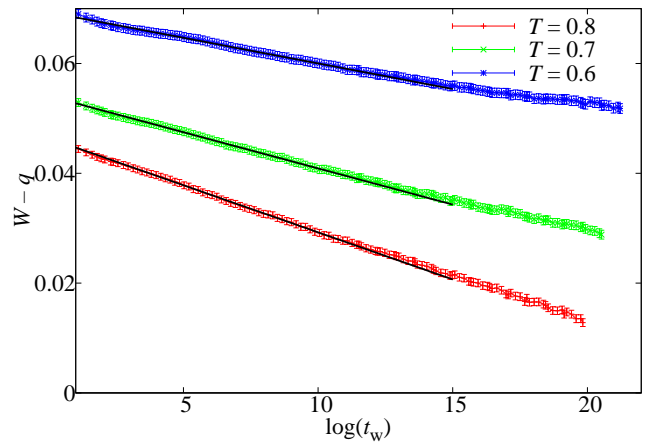}
\end{center}
  \caption{(Color online) 
$W-q$ as a function of $\log(t_\text{w})$ for several low temperatures
and $H=0.2$. Even in this temperature range we can identify a wide linear 
regime.  \label{fig:W-q-log-lowT}}
\end{figure}

To this end we consider an equation analogous to~\eqref{eq:dif_lT}:
\begin{equation}\label{eq:dif_lT_tw}
W(t_\text{w}) - q(t_\text{w}) = a + \frac{b}{t_\text{w}^x}.
\end{equation}
We have computed fits to~\eqref{eq:dif_lT_tw} for all our lower temperatures,
finding that the power-law decay describes the behavior of $W-q$ with great
precision. Indeed we find for the standard figure of merit
$\chi^2/\mathrm{d.o.f.} < 0.5$ (such a small value is due to the strong
data correlation and to the fact that we are computing $\chi^2$ only with the
diagonal part of the covariance matrix). We show in Figure \ref{fig:WqT_tw} the
fit parameters of $x$ and $a$ as a function of temperature (upper and lower
panels respectively). Clearly, this study leads to much lower values of $x$ than
the analysis in $t_\text{tot}$ (but recall that in the previous section 
the value of $x$ decreased if we shifted the fitting window to longer times, 
a problem we do not have here). Indeed, the values of $a$ are more
similar to those computed with~\eqref{eq:WqT_lT_log}. Finally, let us recall
that relaxation exponents of $\mathcal O(10^{-2})$ have already been seen
in the $H=0$ case \cite{janus:09b}.

In any case, the qualitative picture is just the same as in our previous 
section, although the values of $x$ are much lower. This is because, since
we are using the run with the longest $t_\text{base}$, the effective time
at $t_w=0$ is large (in other words, $W-q$ is already very low at $t_w=0$, since
it has already evolved for a considerable time at higher temperatures).

Again, we can use the temperature at which $a$ becomes zero as our estimate of
$T_\text{c}^\mathrm{low}(H)$. We note in Figure~\ref{fig:WqT_tw}--bottom that the
statistical errors for $a$ are small only when it is positive. Hence we have
located the zeros by performing first a linear fit to these points, and then
finding the root of the linear function. In order to take care of the extreme
data correlation, we use a jackknife procedure: fit with the diagonal part of
the covariance matrix, but then perform separate fits for each jack-knife
block \cite{janus:09b}. We obtain $T_\text{c}^\mathrm{low}(H=0.3)= 0.48(2)$,
$T_\text{c}^\mathrm{low}(H=0.2)= 0.69(1)$, $T_\text{c}^\mathrm{low}(H=0.1)=0.906(6)$.

Similarly, we can extrapolate $q(t_\text{w})$ and $W(t_\text{w})$ to infinite
time separately. Again, we obtain $\chi^2/\text{d.o.f.} <1$ in all cases for $t_\text{w}\gtrsim1000$.
We do not reproduce the resulting picture, since it is essentially the same
as Figure~\ref{fig:qT} (with a slightly higher value for $q_\text{min}$ at low $T$).

In short, we can say that assuming a power-law decay of $W-q$ at low
temperatures leads to a picture consistent with a RSB spin-glass transition at
$T_\text{c}^\mathrm{low}(H)\approx T_\text{c}^\mathrm{high}(H)$.

\subsection{Assuming no phase transition: linear-log  extrapolations to large times}~\label{sec:NON-EQUILIBRIUM-LINEAR-LOG}
In the previous section we assumed that the decay of $W-q$ followed a power law
at low temperatures and tried to determine the point where the asymptote became positive.
In this section we take the opposite approach and will assume that there is no phase 
transition, that is, that $W-q$ goes to zero for all $T>0$.

In order to do that, we are going to recall our phenomenological
expression~\eqref{eq:logtau}, which, at high temperature, described 
a wide time range where $W-q$ was linear in $\log(t_\text{w})$.
In Section~\ref{sec:TAUS} we used this functional form
to estimate a characteristic time scale $\tau''$. Naturally, 
the real curve $W-q$ must deviate from~\eqref{eq:logtau}, 
otherwise it would become negative for $t>\tau''$, but 
at high temperature we found that the curvature was noticeable
only at the very end of the simulation, where $W-q$ was already very 
small (even compatible with zero).
\begin{table}[t]
\begin{ruledtabular}
\begin{tabular}{cccc}
$H$ & $T$ & $\log(\tau'')$ & $T \log(\tau'')$\\
\hline
\multirow{5}{*}{0.3} & 0.7 & 27.9(4)  & 19.55(26) \\
                     & 0.6 & 43.0(7)  & 25.8(4) \\
                     & 0.5 & 80.2(1.3)& 40.1(6) \\
                     & 0.4 & 179(4)   & 71.5(1.5) \\
\hline
\multirow{5}{*}{0.2} & 0.8 & 28.4(6)  & 22.7(4)  \\
                     & 0.7 & 42.6(9)  & 29.8(6) \\
                     & 0.6 & 78.8(1.8)& 47.3(1.1)  \\
                     & 0.5 & 178(6)   & 89.1(2.8)\\
                     & 0.4 & 409(13)  & 163(6) \\
\hline
\multirow{7}{*}{0.1} & 1.0 & 26.7(6)  & 26.7(6)    \\
                     & 0.9 & 39.9(9)  & 35.9(8) \\
                     & 0.8 & 67.6(1.6)& 54.1(1.3) \\
                     & 0.7 & 115(3)   & 80.6(2.3)  \\
                     & 0.6 & 226(6)   & 135(4)  \\
                     & 0.5 & 450(14)  & 225(8)\\
                     & 0.4 & ---      &  --- \\
\end{tabular}
\end{ruledtabular}
\caption{Computation of a lower bound for the relaxation time
at low temperature using the linear fit in $\log(t_\text{w})$ 
of Eq.~\eqref{eq:logtau}. For each $H$ we include 
the values of $\log(\tau'')$ for temperatures in the 
non-equilibrium regime (i.e., lower than those 
included in Table~\ref{tab:tau}). For $H=0.1$ and $T=0.4$ 
the curve $W-q$ is flat within errors and we cannot determine 
any $\log(\tau'')$. Data from the cold annealings.
\label{tab:logtau}}
\end{table}

In this section, we are going to use~\eqref{eq:logtau} in order 
to obtain a lower bound for the relaxation time of the system.
Indeed, if we look at Figure~\ref{fig:W-q-log-lowT}, we can 
see that even for very low $T$ the difference $W-q$ 
behaves linearly in $\log(t_\text{w})$ for a long time scale, 
before slowing down its decay. Therefore, in the very reasonable 
assumption that there is no convexity change, we 
can use the parameter $\log(\tau'')$ of the linear fit
in order to obtain a lower bound for the actual 
relaxation time $\tau$ of the system. 

With this procedure, we obtain a finite lower bound
for $\tau$ even for very low temperatures (see Table~\ref{tab:logtau}).
The resulting values of $\log(\tau'')$ are enormous (as an amusing 
comparison, the age $\mathcal T$ of the universe  measured
in MCS is $\log(\mathcal T) \sim 68$, much smaller than some 
of the measured $\tau''$). Therefore, even a rather loose
lower bound gives us  a wildly growing time scale.

In order to make this statement more quantitative, let us recall 
that, in the droplet picture, $\xi(H,T)$ is finite even 
at $T=0$ for $H>0$. Therefore, for low temperatures we would expect
\begin{equation}\label{eq:droplet-lowT}
\log \tau = \xi(0,H)^\Psi/T,
\end{equation}
or, in other words, we would expect $T \log(\tau)$ 
to be constant. However, see Table~\ref{tab:logtau}, 
we find that even our lower bound $\log(\tau'')$ grows 
much faster than predicted by the droplet theory.

\begin{figure}[!t]
\begin{center}
\includegraphics[height=\columnwidth,angle=270]{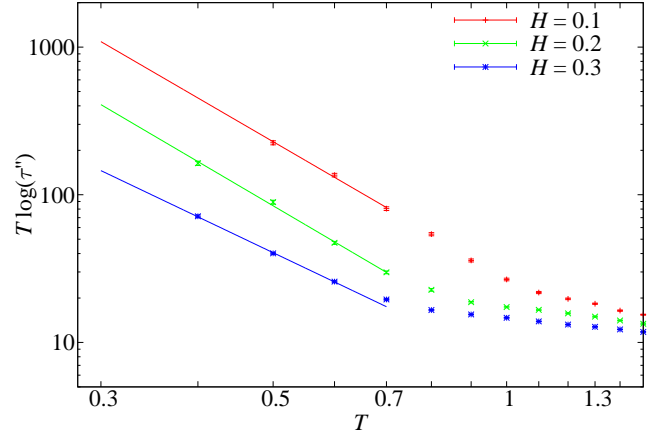}
\end{center}
  \caption{(Color online) 
Plot of a lower bound for $T\log(\tau)$, computed using~\eqref{eq:logtau},
against $T$ for our three magnetic fields. At very low temperatures 
this quantity is compatible with a Vogel-Fulcher-Tammann divergence 
at $T=0$, Eq.~\eqref{eq:VFT-logtau}. Notice 
the change of regime at around $T_\text{c}^\mathrm{high}(H)$ (recall Figure~\ref{fig:T_tau}).
  \label{fig:VFT-logtau}}
\end{figure}

We can take this one step further.  In Figure~\ref{fig:VFT-logtau}
we have plotted $T\log(\tau'')$ against $T$ 
in a log-log scale. As we can see, the data for low temperature
are well described by a Vogel-Fulcher-Tammann divergence 
at $T=0$:
\begin{equation}\label{eq:VFT-logtau-Tc-zero}
T \log(\tau'') = \frac{A}{T^c}\ .
\end{equation}
A fit to~\eqref{eq:VFT-logtau-Tc-zero} for our three magnetic fields 
gives:
\begin{itemize}
\item $H=0.1$, $c(H=0.1)= 3.05(10)$, fitting in $T\leq 0.7$, with $\chi^2/\text{d.o.f.}=2.1/1$.
\item $H=0.2$, $c(H=0.2)= 3.09(6)$, fitting in $T\leq 0.7$, with $\chi^2/\text{d.o.f.}=4.1/2$.
\item $H=0.3$, $c(H=0.2)= 2.50(5)$, fitting in $T\leq 0.6$, with $\chi^2/\text{d.o.f.}=0.80/1$.
\end{itemize}
We can see that $H=0.1$ and $H=0.2$ even have the same exponent, while
$\log(\tau'')$ grows a little more slowly for $H=0.3$  (this is probably
because we have not reached low enough temperatures at $H=0.3$).

Actually, the data in Figure~\ref{fig:VFT-logtau} admit fits of a more general form
\begin{equation}\label{eq:VFT-logtau}
T \log(\tau'') = \frac{A}{(T-T_*)^c}\ ,
\end{equation}
where $|T_*|$ is very small but $T_*$ could even be negative.  Unfortunately, we do
not have enough degrees of freedom to fit simultaneously for $c$ and $T_*$. We shall
discuss the possible implications of this $T_*$ in the following section.

Notice, finally, that in Figure~\ref{fig:VFT-logtau} we can appreciate a sharp
change of regime precisely around the temperature where we identified 
a power-law divergence of $\tau$, fitting from the high-temperature phase.

In the following section we shall introduce a more
direct study of $\xi$ and try to combine the results of Sections~\ref{sec:equilibrium}
and~\ref{sec:NON-EQUILIBRIUM} in a consistent physical picture.

\begin{figure}[t]
\begin{center}
  \includegraphics[height=\columnwidth,angle=270]{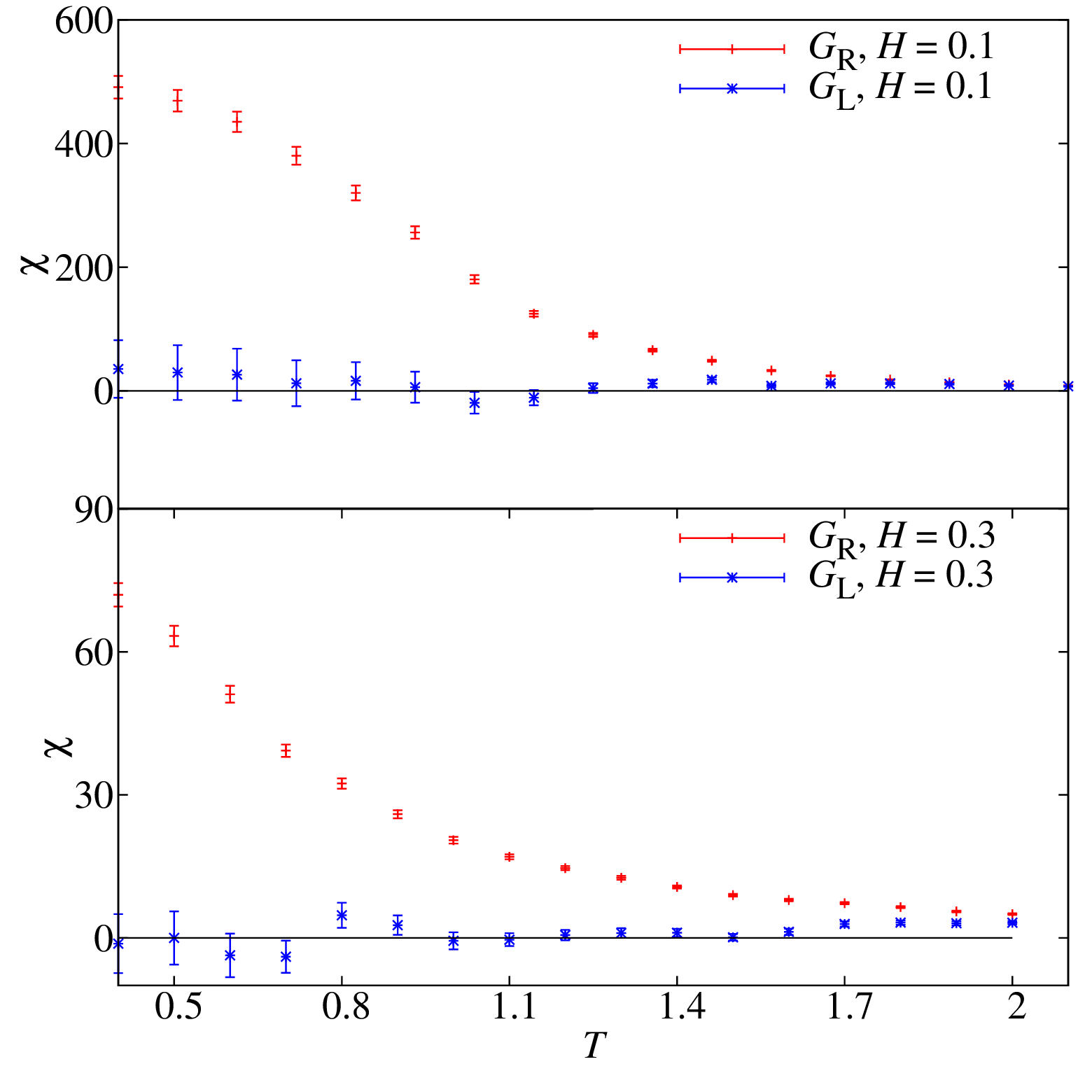}
\end{center}
  \caption{(Color online) Replicon and longitudinal susceptibilities in our 
cold annealings for $H=0.1,0.3$ (we plot the last $t_\text{w}$ for 
each $T$). The plot contains both equilibrium and non-equilibrium
data.
  \label{fig:ji_R}}
\end{figure}

%%%%%%%%%%% XI
\section{Dynamics and the correlation length}\label{sec:xi}

Up to now, we have focused on the determination of characteristic times and
their temperature dependence. However, a proper discussion of any phase
transition requires as well the consideration of spatial correlation and the
correlation length (this is, in fact, a longstanding obstacle in the
investigation of structural
glasses \cite{weeks:00,berthier:05,montanari:06}). In the framework of spin
glasses we are advantaged, because the structure of correlators has been
investigated in detail (see Section~\ref{sec:OBSERVABLES} and references
therein). In particular, there are two types of correlation functions to deal
with, the replicon and the longitudinal/anomalous correlator. We shall first
decide which of the two correlators is worth studying and check that
equilibrium results can indeed be obtained in some temperature range
(Section~\ref{sec:corr-funct}). At that point, we shall revisit the the
dynamics on the view of the correlation length (Section~\ref{sec:xi_T_tw}).

\subsection{Which correlation function?}\label{sec:corr-funct}
We compare in Figure~\ref{fig:ji_R} the replicon and longitudinal/anomalous
susceptibilities [recall Eq.~\eqref{eq:chi_def}], for $H=0.1, 0.2$ and $0.3$.
The susceptibilities are shown as a function of temperature, as computed for
the latest time on each temperature step along the cold annealing. This means
that Figure~\ref{fig:ji_R} contains both equilibrium and non-equilibrium data
(depending on whether the constant-temperature step is much larger than $\tau$,
or not). In either case, it is rather obvious that significant correlations
appear only on the replicon correlator (in agreement with equilibrium,
mean-field computations \cite{dealmeida:78}). Therefore, we focus
on the replicon correlator from now on. The anomalous sector is studied in more
detail in Appendix~\ref{app_GRGL}, Section~\ref{app_GRGL:anticorrelations}.

  We note as well that the failure of
the longitudinal/anomalous correlator to display the correlations relevant to
the problem might be related to analogous failures in experimental
investigations of structural glasses \cite{debenedetti:97,debenedetti:01}.
\begin{figure}[t]
\begin{center}
  \includegraphics[height=\columnwidth,angle=270]{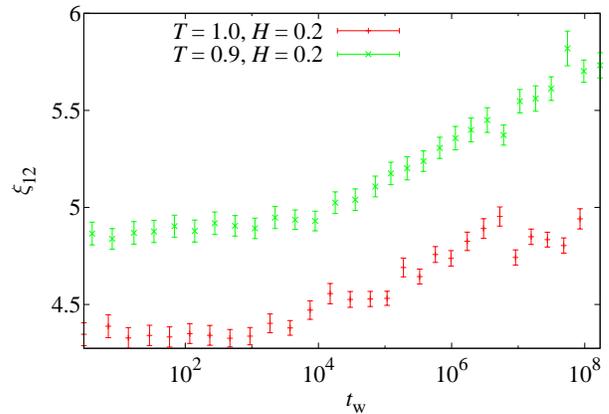}
\end{center}
  \caption{(Color online) Equilibration of the correlation length: for $H=0.2$, we show
the $\xi_{12}$ correlation length as
computed from the replicon correlator $G_\text{R}$ as a function of $t_\text{w}$. We display
data for two constant-temperature steps in the
cold annealing (recall that $t_\text{w}$ is the time elapsed since the last
temperature drop). The constant temperature time step is longer than $\tau$
only for $T=1.0$. }
  \label{fig:xi_log_tw}
\end{figure}

Specifically, we consider the $\xi_{12}$ correlation length as
computed from the replicon correlator $G_\text{R}$ using Eq.~\eqref{eq:xi_integrals}.
Its time evolution for two constant-temperature steps in the
cold annealing run is displayed in Figure~\ref{fig:xi_log_tw}. For both
temperatures we identify three regimes. For short $t_\text{w}$ the correlation length
basically remains constant (the fact that $\xi_{12}$ does not decrease at
$t_\text{w}\sim 1$ is interesting in itself: it tells us that temperature chaos
effects are weak). Then, the time evolution starts to be noticeable and $\xi$
starts to increase. Finally, when $t_\text{w}\gg \tau$, the correlation length becomes
time-independent, which is consistent with our physical interpretation in
Section~\ref{sec:equilibrium} that thermal equilibrium has been reached. We note
as well that the equilibrium regime is barely reachable for $T=0.9$ ---remember
that $\tau''(T=0.9)=1.1(3)\times 10^9$, while $\tau''(T=1.0)=1.8(7)\times10^7$.
In the next paragraph, we shall discuss $\xi_{12}$ as
a function of temperature, but only for those temperatures where thermal
equilibrium can be reached.

\subsection{Dynamics from the point of view of the correlation length}\label{sec:xi_T_tw}

An important question is: is the dynamics activated [
  $\tau\sim\mathrm{e}^{\xi^\varPsi/T}$, Eq.~\eqref{eq:droplets-dynamics}], or
critical [$\tau\sim \xi^z$, Eq.~\eqref{eq:critical-dynamics}]?

As explained in Section~\ref{sec:droplets_vs_RSB}, the droplet theory supports
activated dynamics. On the other hand, the RSB theory is somewhat ambiguous on
this point [at mean-field level the dynamics \emph{is} critical, this is the
  rationale for using $z\nu$ as the critical exponent in
  Eq.~\eqref{eq:div_tau}]. Furthermore, current theories for supercooled
liquid relaxations predict \emph{both} types of behaviors
(Section~\ref{sec:MCT-and-beyond}). These theories expect critical dynamics at
high temperatures, with an effective exponent $z_\mathrm{MCT}=4 \gamma$ [recall
  Eqs.~(\ref{eq:MCT-1},\ref{eq:MCT-2})]. However, at lower temperatures the
dynamics should crossover to an activated behavior.

\begin{figure}[t]
\begin{center}
  \includegraphics[height=\columnwidth,angle=270]{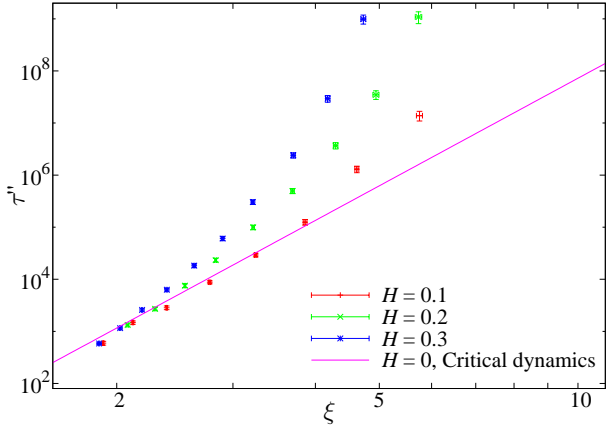}
\end{center}
  \caption{(Color online) Logarithmic plot of $\tau''$, computed in Section~\ref{sec:PROTOCOLS_AND_TAUS},
       versus the correlation length ($\xi_{12}$) for
    the three magnetic fields simulated. Data are in equilibrium. For comparison, we also show the
    critical dynamics for $H=0$, $\tau \sim \xi^z$, with
    $z=6.86$ \cite{janus:09b}. \label{fig:xi_tau}}
\end{figure}

At this point, we have in our hands equilibrium determinations for both
$\tau(T,H)$ and $\xi_{12}(T,H)$. Therefore, we can try to assess
Eqs.~(\ref{eq:droplets-dynamics},\ref{eq:critical-dynamics}) directly.  This
is attempted in Figure~\ref{fig:xi_tau}, where we used $\tau''$ from fits to Eq.~\eqref{eq:logtau}.
Although this choice is arbitrary to some extent, we
recall that the critical divergence studied in Section~\ref{sec:equilibrium}
turned out to be independent on the choice of $\tau$. 

We note two different regimes in Figure~\ref{fig:xi_tau}. For high-temperatures,
data follow a critical dynamics. However, at lower temperatures (i.e., larger
$\xi$) $\tau$ starts to grow much faster with $\xi$. In fact, the effective
exponent $z^{\text{eff}}=\mathrm{d\, log\,}\xi/\mathrm{d\, log\, \tau}$
becomes as large as $z^{\text{eff}}\approx 14$, which
clearly suggest that the dynamics is becoming activated. Overall, this
crossover exemplifies the behavior expected for a supercooled liquid. In fact,
critical dynamics is found in the range $10^2 < \tau < 10^4$, which is also
the range where MCT applies for simple supercooled liquids \cite{kob:94}.
However, an alternative interpretation is possible. First, one may note that
we identified in Section~\ref{sec:tau-mct} an
exponent $\gamma\approx 2$. Considering the large uncertainty 
in this determination, this is consistent, via Eqs.~\eqref{eq:MCT-1} and~\eqref{eq:MCT-2}, 
with the $z_\text{eff} \approx 7$ that we find in Figure~\ref{fig:xi_tau}.
However, the slope in the figure is also very close to $z_{H=0}=6.86$, the value for the
critical dynamics in the absence of a magnetic field \cite{janus:09b}. Hence,
the crossover can be also due to the proximity of the Renormalization-Group
fixed point at $(T_\mathrm{c},H=0)$. In fact, the larger $H$ is the
smaller the $\xi$ needed to find activated dynamics (see
Figure~\ref{fig:xi_tau}).

\begin{figure}[t]
\begin{center}
  \includegraphics[height=\columnwidth,angle=270]{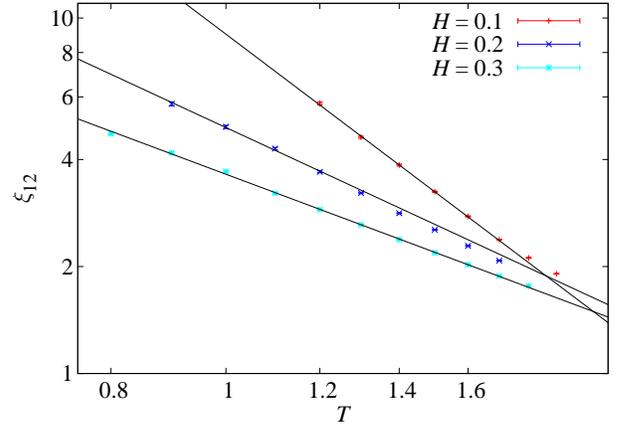}
\end{center}
  \caption{(Color online) Logarithmic plot of the equilibrium correlation length $\xi_{12}$
    versus the temperature for the three magnetic fields simulated. Lines are
    fits to $\xi_{12}(T)=a_H/T^{b_H}$, see
    Eq.~\eqref{eq:fit_xi_T}. When assessing the high-temperature data for
    $H=0.1$, recall that the critical temperature without a field is
    $T_\mathrm{c}^{H=0}=1.1019(29)$ \cite{janus:13b}.}
  \label{fig:xi_T}
\end{figure}

Let us consider the temperature evolution of $\xi_{12}$ in equilibrium, see
Figure~\ref{fig:xi_T}. We do not find good fits to critical divergences,
$\xi_{12}\propto 1/|T-T_\mathrm{c}(H)|^\nu$, with $T_\mathrm{c}(H)$ compatible
with the characteristic temperatures identified in
Section~\ref{sec:equilibrium}.  For instance, a fit for $H=0.2$ gives
a reasonable  $\chi^2$ value only for $T_\text{c} \lesssim 0.5$, while 
at $H=0.1$ we get $T_\text{c}\lesssim0.8$ (these bounds are very crude,
since we have almost no degrees of freedom for the fits). 
In particular, assuming that $T_\text{c}=0$ we still get good fits:
\begin{eqnarray}
\xi_{12}(T)&=&\frac{a_H}{T^{b_H}}\,,\label{eq:fit_xi_T}\\
b_{H=0.1}&=&2.52(4)\quad (1.2\leq T \leq 1.5,  \chi^2/\text{dof}=1.9/2)\,.\\
b_{H=0.2}&=&1.55(4)\quad (0.9\leq T \leq 1.2, \chi^2/\text{dof}=3.6/2)\,.\\
b_{H=0.3}&=&1.10(5)\quad (0.8\leq T \leq 1.0, \chi^2/\text{dof}=0.25/1)\,.
\end{eqnarray}
Hence, at least within the temperature range that can be equilibrated, our
data are compatible with a divergence of $\xi$ only for very low (perhaps even vanishing)
$T_\mathrm{c}(H)$. Notice, however, that we are always working with $\xi<6$, so
this kind of fit is rather dangerous.

\begin{figure}[t]
\begin{center}
  \includegraphics[height=\columnwidth,angle=270]{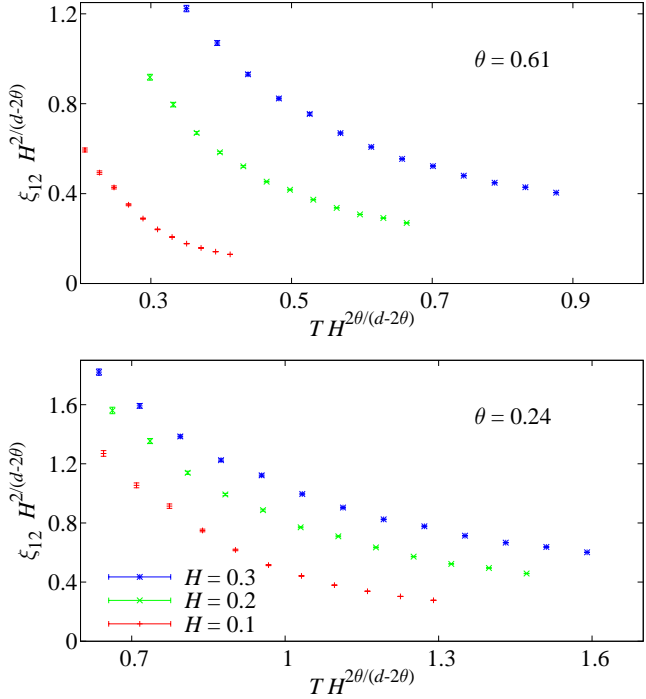}
\end{center}
  \caption{(Color online) Test of the droplet scaling law, Eq.~\eqref{eq:droplets-for-xi}.
    For each of our magnetic fields, we plot $\xi(T,H)\, H^{2/(d-2\theta)}$,
  as a function of $T H^{2\theta/(d-2\theta)}$. For a proper choice of the stiffness
  exponent $\theta$, data should collapse in a single curve. We try (and fail
  within the reachable temperature window)
  to collapse the data with two different estimates of $\theta$, see
  Section~\ref{sec:droplets_vs_RSB}}
  \label{fig:xi-droplets}
\end{figure}

At this point, is is only natural to ask whether the droplet theory, see
Eq.~\eqref{eq:droplets-for-xi}, describes our data. The answer is negative,
see Figure~\ref{fig:xi-droplets}.

Therefore, none of the available theories provide a satisfactory description
of our simulation. Of course, this might be due to the fact that we have not
reached the regime where these theories apply (low enough temperatures, or low
enough magnetic fields). However, we should stress that our data span a rather
significant range of time scales (from one picosecond to a hundredth of a
second). Hence, we dare say that our simulations are of direct experimental
relevance. The issue is discussed at length in the Conclusions.

A final remark. One can be tempted to compare Eq.~\eqref{eq:fit_xi_T}, which
works for our equilibrium data, with the analysis in
Section~\ref{sec:NON-EQUILIBRIUM-LINEAR-LOG} giving $T \mathrm{log}\tau\sim
1/T^c$ (which is based on an extrapolation to times beyond our simulated
timescales). This comparison tells us that the droplet exponent $\varPsi\approx
c/b$. Therefore, our data for $H=0.1$ suggest $\varPsi\approx 1.5$, while our
results for $H=0.2,0.3$ rather suggest $\varPsi\approx 2$. These values are
rather large, as compared to the value $\varPsi\sim 0.03$ found in
\cite{bert:04}.

%%%%%%%%CONCLUSIONS
\section{Conclusions}\label{sec:conclusions}

In this work, we have investigated the approach to equilibrium and the
building-up of spin-glass order for the $d=3$ Ising spin glass in an external
magnetic field. Specifically, we have simulated the Edwards-Anderson model on
the Janus dedicated computer. Our lattices were always much larger ($L=80$)
than the correlation length, $\xi(t_\text{w})$. Hence, we think that our
results are representative of the thermodynamic limit. Our time scales range
from the picosecond to one hundredth of a second. We are thus approaching the
experimental scale. However, when the temperature was low enough, we have been unable
to reach the thermalization time scale, $\tau$. We have  monitored this
effect carefully. Therefore, in this work we are presenting in a controlled way both
equilibrium and non-equilibrium data. Our results have been analyzed on the light
of the two major theories on the market, the replica-symmetry breaking and the
droplet theory. On the view of recent claims \cite{moore:02,fullerton:13}, we
have also analyzed our data as suggested by current theories for
relaxation in supercooled liquids. None of these three approaches was fully
satisfying.

The problem with the droplet/RSB theories was in the correlation length: the
growth of $\xi$ upon lowering the temperature is too fast to fit
the droplet theory and too slow to fit RSB.  We summarize now the strengths and
weaknesses of each approach. We start with the droplet theory, then consider RSB
and, finally, the supercooled liquids point of view.

As we show in Section~\ref{sec:xi}, the dynamics really seems to be of activated
type, as predicted by the droplet theory. However, the scaling law predicted
by the theory is not fulfilled by our data. In fact, see
Section~\ref{sec:NON-EQUILIBRIUM-LINEAR-LOG}, the dynamics is of super-Arrhenius
type at least down to temperature $T=0.4$ (to be compared with
$T_c=1.1019(29)$ \cite{janus:13b}, the $H=0$ critical
temperature). Therefore, although the droplet theory might be finally correct
at still lower temperature, the corresponding time and length scales would be
beyond not only our computational capabilities, but also current 
experimental possibilities.

The RSB approach resulted in a determination of the de Almeida-Thouless line,
which is consistent, whether one uses equilibrium
(Section~\ref{sec:equilibrium}) or non-equilibrium
(\ref{sec:NON-EQUILIBRIUM-POWER-LAW}) data. Unfortunately, our
\emph{equilibrium} estimate of $\xi$ does not seem to diverge at the de
Almeida-Thouless line (also, the Fisher-Sompolinsky scaling\cite{fisher:85} is \emph{not}
verified, as the reader may easily check). There are a number of possible
explanations for our failure to find the divergence:
\begin{itemize}
\item For all three magnetic fields, we have been able to equilibrate the system only down
  to $T\approx 1.3 T_\mathrm{c}(H)$. Perhaps the critical growth
  of $\xi(T,H)$ starts only closer to the de Almeida-Thouless line.
\item It is by no means guaranteed that we are looking at the right
  correlation function. We have shown in Section~\ref{sec:corr-funct} that some
  correlators might display sizeable correlations while others do not. In
  fact, the quest for sensible correlators is a longstanding problem in the
  investigation of supercooled
  liquids \cite{franz:00,kirkpatrick:88,weeks:00,berthier:05,montanari:06,cavagna:07,biroli:08}.
  Also in the field of spin glasses it has been suggested that energy and
  link-overlap correlators deserve more
  attention \cite{marinari:98d,contucci:05b,contucci:06,janus:10}.
\item As explained in Section~\ref{sec:droplets_vs_RSB}, it is very possible
  that the physics in $d=3$ will be ruled by a fixed point at $T=0$. If this
  is the case, activated dynamics is to be expected also in the RSB theory.
  Under these circumstances, the de Almeida-Thouless line identified in
  Section~\ref{sec:equilibrium} and~\ref{sec:NON-EQUILIBRIUM-POWER-LAW} might
  well represent a dynamic glass transition. In fact, our data for $\xi(T,H)$
  are consistent with a critical divergence \emph{below} the de
  Almeida-Thouless line. The divergence could take place in the
  range $0\leq T_\mathrm{c}\leq 0.8$ for $H=0.1$, $0\leq T_\mathrm{c}\leq 0.5$
  for $H=0.2$ and $0\leq T_\mathrm{c}\leq 0.5$ for $H=0.3$ (note, however,
  that these upper bounds are only crude estimations).
\item Finally, as explained above, it is possible that the droplet theory could
be correct and no transition takes place for $H>0$ in three spatial dimensions.
However, it is clear that the correlation length grows noticeably in our
simulations, which suggests that the lower critical dimension in a field cannot
be much larger than $d=3$. Notice as well that the lower critical dimension in
a field is well below $d=4$, since very clear evidence of a
second-order phase transition has been obtained in $d=4$ \cite{janus:12}.
\end{itemize}

Let us finally consider the supercooled liquid  approach. At the
qualitative level, this is maybe the most successful description. Indeed, we do
identify in our data a Mode Coupling temperature, $T_\text{d}$
(Section~\ref{sec:tau-mct}) and a cross-over to activated dynamics
(Sects.~\ref{sec:tau-mct} and~\ref{sec:xi_T_tw}). However, the description
remains qualitative since our numerical accuracy does not allow a strict
test of basic relations among critical exponents. We note as well that the
would-be Mode Coupling temperature for $H=0.2$, $T_\mathrm{d}\approx 1.22$ is
rather large as compared to the de Almeida-Thouless line
$T_\mathrm{c}(H=0.2)\approx 0.7$ (see Section~\ref{sec:equilibrium}). In this
respect, we remark that a more typical value for supercooled liquids is
$T_\mathrm{d}\approx 1.1 T_\mathrm{g}$ ($T_\mathrm{g}$ is the dynamic glass
temperature where $\tau\sim$ 1 hour).

We conclude by mentioning that a further finite-size scaling investigation of
the problem is now ongoing.

\section*{Acknowledgments}

The Janus project has been partially supported  by the  EU (FEDER funds, No.
UNZA05-33-003, MEC-DGA, Spain); by the European Research
Council under the European Union's Seventh Framework Programme (FP7/2007-2013,
ERC grant agreement no.  247328); by the MICINN (Spain) (contracts FIS2006-08533,
FIS2012-35719-C02, FIS2010-16587, TEC2010-19207);
by the SUMA project of INFN (Italy); by CAM (Spain); by the Junta de
Extremadura (GR10158); by the Microsoft Prize 2007 and by the European Union (PIRSES-GA-2011-295302).
F.R.-T. was supported by the Italian Research Ministry through the
FIRB project No. RBFR086NN1; M.B.-J. was supported by the FPU program (Ministerio de Educacion, Spain);
R.A.B. and J.M.-G. were supported by the FPI program (Diputacion de Aragon,
Spain); S.P.-G. was supported by the ARAID foundation; finally J.M.G.-N. was
supported by the FPI program (Ministerio de Ciencia e Innovacion, Spain). 

\appendix

\section{Discretization of the Gaussian Magnetic Field}
\label{app1}

\begin{figure}[t]
\includegraphics[height=\columnwidth, angle=270]{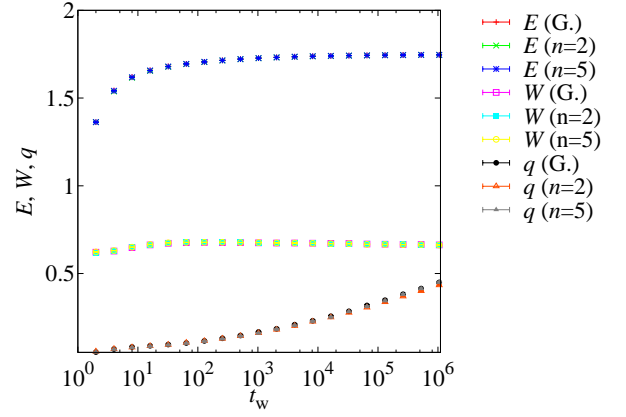}
  \caption{(Color online) Thermal energy $E(t_\text{w})$, magnetic energy
$W(t_\text{w})$ and overlap $q(t_\text{w})$ as a function of time for $L=8$,
$T=0.7$ and $H=0.3$. We have plotted the results from a full Gaussian
distribution (G.), as well as discretizations with  $n=2$ and $n=5$. Notice
that all three simulations have produced the same equilibrium values 
for these observables.}
  \label{fig:test_8}
\end{figure}

In this appendix we describe the procedure we have used to discretize (using
the Gauss-Hermite quadrature~\cite{abramowitz:72}) the
Gaussian magnetic field in order to implement our simulations on
the Janus dedicated computer (which cannot handle non-integer
arithmetic efficiently). This implementation was introduced (but not 
explained in detail) in \cite{leuzzi:09}.

The Gauss-Hermite quadrature formula can be interpreted as an approximation
formula for \emph{probability distributions}.  In fact, if we multiply either
of the two distributions related in Eq.~\eqref{eq:gauss-hermite}, below, by an
arbitrary polynomial of order $4 n-1$, and integrate this product through
$x\in (-\infty,\infty)$, identical results are obtained \cite{abramowitz:72}:
\begin{equation}\label{eq:gauss-hermite}
e^{-x^2} \approx
\sum_{k=1}^n\, \frac{w_k}{2} [\delta(x -x_k) + \delta(x+x_k)]\,.
\end{equation}
where $x_k$ are the positive zeros of the $2n$-th Hermite polynomial $H_{2n}(x)$ and the
weights, $w_k$, are given by 
\begin{equation}
\label{eq:weights}
w_k=\frac{2^{2n-1} (2n)! \sqrt{\pi}}{(2n)^2 H_{2n-1}(x_k)^2} \, .
\end{equation}

In our numerical simulations we need to compute integrals like
\begin{equation}
\overline{O}\equiv \frac{1}{H \sqrt{2 \pi}}
\int_{-\infty}^\infty dh~  O(h) e^{-\frac{h^2}{2 H^2}} 
\end{equation}
Furthermore, the above equation can be further simplified using the gauge
symmetry~\eqref{eq:gauge-symm}, which allows us to consider only positive
magnetic fields, so
\begin{equation}
\overline{O}=\frac{2}{H \sqrt{2 \pi}}
\int_{0}^\infty dh~  O(h) e^{-\frac{h^2}{2 H^2}} 
\end{equation}
 and using Eq.~\eqref{eq:gauss-hermite}, one can finally write
\begin{equation}
\overline{O}=\frac{2}{\sqrt{\pi}} \sum_{k=1}^n w_k O(\sqrt{2} H x_k) \,.
\end{equation}
We shall limit ourselves to $n=2$ in Eq.~\eqref{eq:gauss-hermite}. Hence, the
magnetic field for each site of the lattice is chosen independently: with
probability $\tilde w_1=2 w_1/\sqrt{\pi}$ the field is $h_1=\sqrt{2} H x_1$
(it is set to $ h_2=\sqrt{2} H x_2$ otherwise). Note that two bits per lattice
site are enough to code this $n=2$ approximation, which is very important
given the limited memory in Janus. The Gauss-Hermite values for $n=2$ are
$x_1=0.524647623275$, $x_2= 1.65068012389$, $w_1=0.804914090006$ and
$w_2=0.0813128354472$ [see \cite{abramowitz:72} or use
Eq.~\eqref{eq:weights}].

We have tested numerically the accuracy of the $n=2$-approximation by
performing some numerical tests on a small lattice size ($L=8$).  Obviously,
our $n=2$ approximation should fail for high magnetic fields and small lattice
sizes. We have checked that our choice is valid at least for $H \le 0.3$. In
particular, we have compared the results of simulations with $n=2$ (our choice
in this work), and $n=5$ with the full Gaussian distribution for the following
observables:  energy,
overlap and $W$(see Figure~\ref{fig:test_8}). We have checked that the
differences between the observables computed at finite $n$ and those computed
with full Gaussian magnetic field are statistically compatible with zero.

Another strong test of our implementation is the agreement in the asymptotic
values of $q(t_\text{w})$ and $W(t_\text{w})$ in the high-temperature phase
(see Figure~\ref{fig:W-q}). 

In conclusion: we have checked that, for the main quantities considered in
this work (computed with $n=2$), the systematic error in the
approximation~\eqref{eq:gauss-hermite} is smaller than our statistical
accuracy.

%%%%%%%%%%APPENDIX B
\section{Spatial correlation functions}
\label{app_GRGL}

In this appendix we discuss three different features of the spatial
correlation functions of the $d=3$ spin glass in a magnetic field: (i) the
long distance behavior of the replicon correlator (recall
Section~\ref{sec:OBSERVABLES}), (ii) the non-monotonic time behavior in
\emph{direct quenches} (this anomaly seems to be absent from annealing
protocols), and (iii) the comparison of the replicon and the
longitudinal/anomalous correlators.
 
\subsection{Long-distance behavior of the replicon propagator}\label{app_GRGL:truncacion}
\begin{figure}[t]
\includegraphics[height=\columnwidth, angle=270]{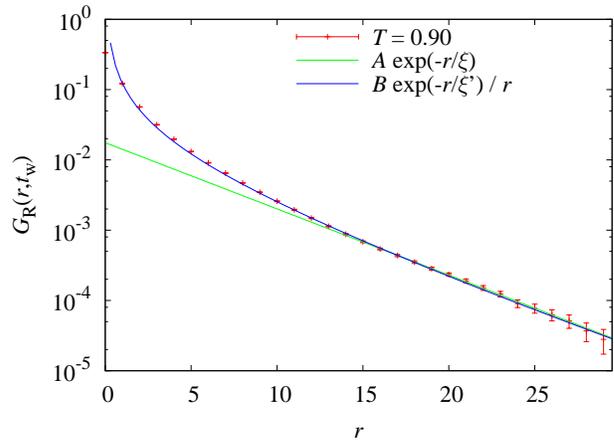}
  \caption{(Color online) Semilogarithmic plot of the equilibrium replicon
    correlator $G_\text{R}$ as a function of distance $r$, for temperature $T=0.9$ and $H=0.2$
    [correlations along the lattice direction $(r,0,0)$]. The lines are fits to
    a single exponential, $G_\text{R}(r)\approx A \mathrm{e}^{-r/\xi}$ ($A$ and $\xi$
    are fitting parameters), obtained for $r\geq 15$ and to $G_\text{R}\approx B \mathrm{e}^{-r/\xi'}/r$ for $r\geq 10$.
   These two functional forms are indistinguishable for large $r$.
  \label{fig:exponential_decay}}
\end{figure}

As explained in Section~\ref{sec:OBSERVABLES}, when computing the integrals
$I_k$, see Eq.~\eqref{eq:integrals}, it is crucial to impose a long distance
cutoff. Otherwise, the $I_k$ integrals become non-self-averaging objects that
can be accurately computed only with a huge number of samples. However, one
may try to correct the systematic effects induced by the cutoff by studying
the long-distance behavior of the propagator $G_\text{R}$. One fits the curve to a
suitable, simple functional form and then computes by hand the remaining part
of the integral. The contribution to $I_k$ from $r>r_\text{cutoff}$ is
usually tiny, but we prefer to monitor it. This issue has been discussed at
length in \cite{janus:08b,janus:09b}, where the spin glass without
a field was studied. 

Here, we show in Figure~\ref{fig:exponential_decay} that, in the temperature
regime where we manage to equilibrate the system in a field, $G_\text{R}(r\gg\xi)$
decays exponentially. Therefore, estimating the contribution from
$r>r_\text{cutoff}$ to the integrals $I_k$ is fairly easy (the resulting correction
is smaller than the error bar). We can include an algebraic prefactor in the fitting
function the better to fit
the small-$r$ sector, but this is irrelevant for the tails (see Figure~\ref{fig:exponential_decay}).

\subsection{Overshooting in the direct quench}\label{app_GRGL:time}

The direct quench is an idealized temperature-variation protocol: one takes a
fully disordered spin glass (i.e.. $T=\infty$) and places it
\emph{instantaneously} at the working temperature. It is clear that, in the
laboratory, temperature should vary gradually. Hence, the annealing protocols
described in Section~\ref{sec:PROTOCOLS} are closer to the
temperature variations that one can realize experimentally. On the other hand,
the direct quench is the simplest protocol in a computer simulation.

\begin{figure}[t]
\includegraphics[height=\columnwidth, angle=270]{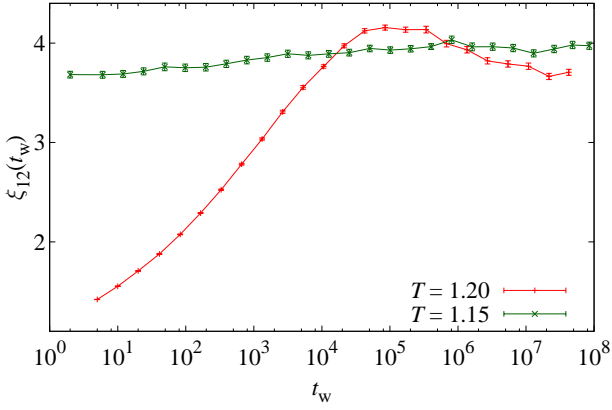}
  \caption{(Color online) Time evolution of the correlation length $\xi_{12}$
    as computed for the replicon correlator for temperatures $T=1.2$ and
    $T=1.15$ in our hot annealing, see Section~\ref{sec:PROTOCOLS}. Note that
    the temperature step $T=1.2$ can be regarded as a direct quench, while
    $T=1.15$ already belongs to the annealing part of the run. The maximum for
    $T=1.2$, where $\xi_{12}(t_\text{w})$ is larger than its equilibrium
    value, seems to be a generic feature of any direct quench for a spin glass
    in a field in three dimensions. On the other hand, during the annealing,
    the time evolution of $\xi_{12}(t_\text{w})$ is monotonic.  }
  \label{fig:overshooting-xi}
\end{figure}
In fact, we have found with some surprise that the non-equilibrium behavior of
the replicon correlator is rather different in a direct quench and in an
annealing protocol. In Figure~\ref{fig:overshooting-xi} we show the time
evolution of the correlation length $\xi_{12}(t_\text{w})$ for two
temperatures in our hot annealing, the initial one $T=1.2$, and the second
temperature $T=1.15$. Note that the time evolution at the very first
temperature in the annealing can be aptly described as a direct
quench. Indeed, in Figure~\ref{fig:overshooting-xi} we notice an overshooting of
$\xi(t_\text{w})$ in the direct quench: well before equilibrium is reached, a
maximum is found which lies above the equilibrium correlation length. No such
maximum arises in the lower temperatures of the annealing. We have checked that
this overshooting is characteristic of the direct quench, as it happens
basically for all temperatures and magnetic fields.

\begin{figure}[t]
\includegraphics[height=\columnwidth, angle=270]{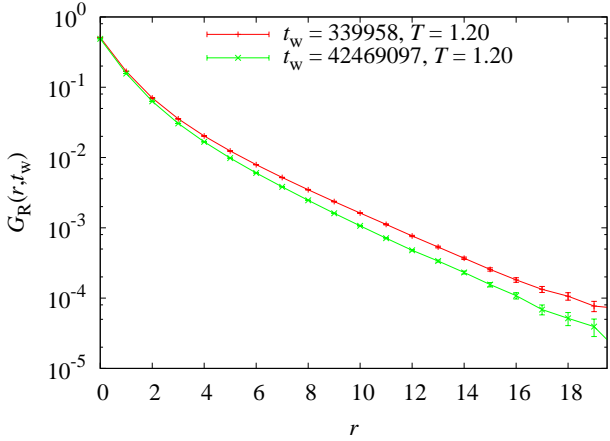}
  \caption{(Color online) Comparison of the non-equilibrium correlator
$G_\text{R}(r;t_\text{w})$ with the corresponding equilibrium value for the initial
  temperature, $T=1.2$, in our hot annealing (i.e., in a direct quench
  run). The time $t_\text{w}$ corresponds to the maximum correlation length in
  Figure~\ref{fig:overshooting-xi}.}
  \label{fig:overshooting-GR} 
\end{figure}
We can look at this overshooting in greater detail in
Figure~\ref{fig:overshooting-GR}, where we compare the equilibrium $G_\text{R}(r)$ with
the non-equilibrium $G_\text{R}(r;t_\text{w})$ at the $t_\text{w}$ corresponding to the
maximum in Figure~\ref{fig:overshooting-xi}. The two correlators are remarkably
featureless as a function of $r$, but the overshooting effect is also clear
from $G_\text{R}(r)$.

\subsection{The anomalous/longitudinal sector}\label{app_GRGL:anticorrelations}

\begin{figure*}[t]
\includegraphics[height=0.8\textwidth, angle=270, trim=250 40 0 40]{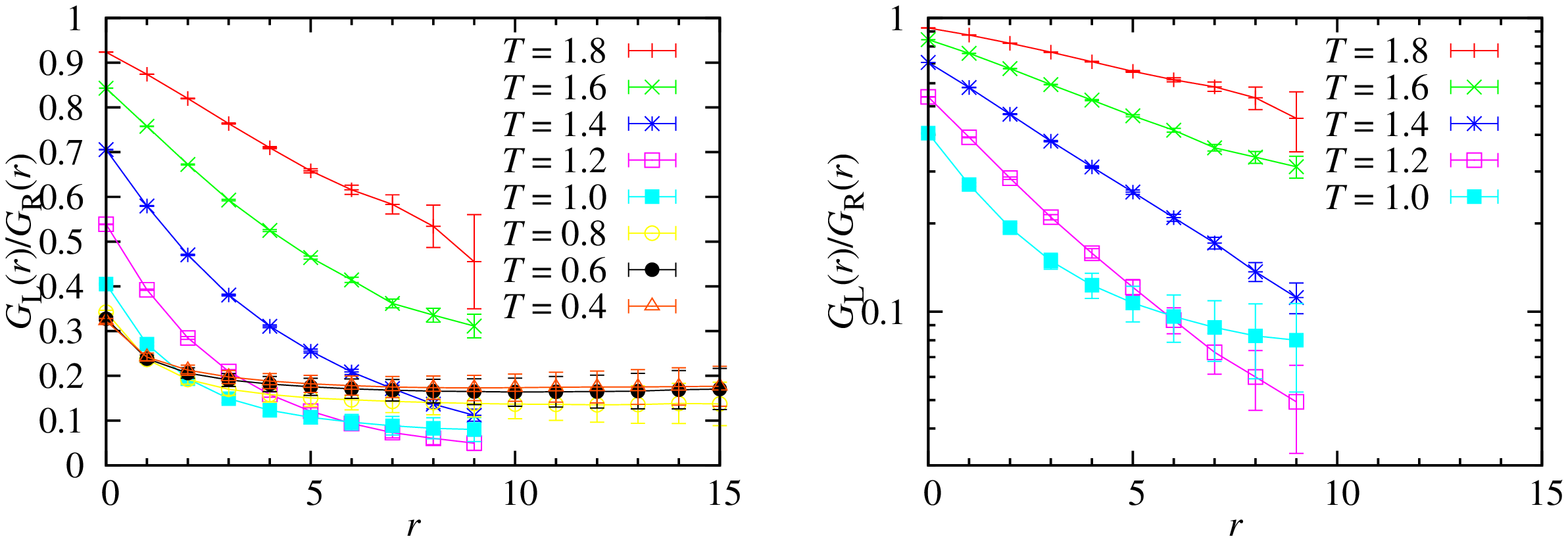}
  \caption{(Color online) 
Comparison of the replicon and the longitudinal
    propagator for $H=0.1$. Data from
    the longest $t_\text{w}$ at each temperature step, in our 
    slowest annealing for each temperature. Notational conventions are as in
    Figure~\ref{fig:exponential_decay}. The left panel shows the quotient $G_\text{L}/G_\text{R}$
    for the whole temperature range in a linear scale, while the right panel shows the same 
    quantity in a logarithmic scale (removing the lowest temperatures). In all cases we cut each graph 
    at the point where the statistical error  becomes greater than $50\%$. For low $T$, the quotient seems
    to reach a non-zero asymptotic value. For high temperatures, it seems to decay exponentially, although
    the decay slows down for large $r$, suggesting a possible non-zero plateau for large $r$.
     \label{fig:GL_GR}}

\includegraphics[height=0.8\textwidth, angle=270, trim=250 40 0 40]{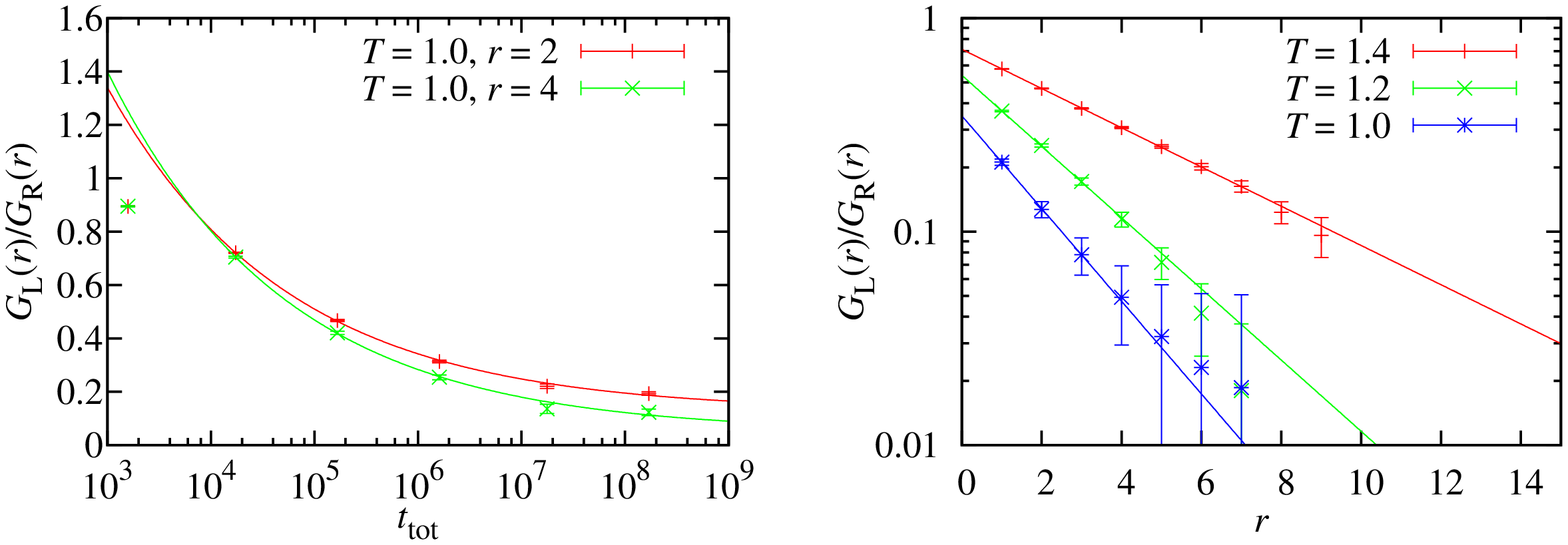}
  \caption{(Color online) \emph{Left:} We plot the quotient $G_\text{L}/G_\text{R}$
  for $T=1.0$ and $H=0.1$ as a function of the total time $t_\text{tot}$ for two 
  different values of $r$. It is clear that even our longest times have small thermalisation effects.
  The long-$t_\text{tot}$ limit may be estimated with a power-law fit (continuous lines). 
   \emph{Right:} $G_\text{L}/G_\text{R}$, extrapolated to infinite $t_\text{tot}$ for several
   temperatures. The curvature observed in Figure~\ref{fig:GL_GR} has disappeared and 
  a pure exponential decay is now apparent.
\label{fig:GL_GR2}}
\end{figure*}

The longitudinal/anomalous correlator defined in Eq.~\eqref{eq:GLGA_n_to_zero}
appears naturally in the analysis of the mean-field
approximation \cite{dealmeida:78}. To the best of our knowledge, the
longitudinal/anomalous correlator has not been studied in three spatial
dimensions, in the presence of a field. We recall that from these correlators,
one may obtain associated susceptibilities, see Eq.~\eqref{eq:chi_def}.

We showed in Figure~\ref{fig:ji_R} that the replicon susceptibility
grows significantly upon lowering the temperature, while the
longitudinal/anomalous susceptibility does not. However, when looking at the plot of $\chi$,
which is a spatial integral of $G$, we are losing information 
on the shape of the correlation function.

Here we perform a more detailed comparison of both correlators by studying
their ratio as a function of $r$ in Fig.~\ref{fig:GL_GR},
left. $G_\text{L}/G_\text{R}$ was computed at the longest time available,
i.e., as close as possible to thermal equilibrium.  We identify two different
regimes, at high and low temperatures. At high temperatures,
$G_\text{L}/G_\text{R}$ decreases exponentially in $r$, see the right panel in
Fig.~\ref{fig:GL_GR}. In fact, barring unavoidable differences on the
algebraic prefactors, $G_\text{L}(r)/G_\text{R}(r)\propto \text{exp}\big[-r
  \,\big(\frac{1}{\xi_\text{L}}-\frac{1}{\xi_\text{R}}\big)\big]$. Hence the
exponential decrease in Fig.~\ref{fig:GL_GR}--right implies
$\xi_\text{R}>\xi_\text{L}$. On the other hand, at the lowest temperatures
that we reach, i.e.,  $T=0.4$ and $0.6$, $G_\text{L}/G_\text{R}$ becomes
essentially constant at large $r$, suggesting that the correlation length is
the same for both correlators. This is quite surprising: if a de
Almeida-Thouless line exists one expects $0 < \xi_\text{L}/\xi_\text{R}<
1$ as we approach it. There is no obvious reason for the two correlation
lengths to be equal.

However, the above could be too hasty a conclusion. The reader might be
surprised (as we were) by the non-monotonic temperature behavior in
the left panel of Fig.~\ref{fig:GL_GR}. One may note that we are
mixing thermalized and non-equilibrium data in that figure, which may
confuse the situation. An example of the time evolution is shown in
the left-panel of Fig.~\ref{fig:GL_GR2}. Clearly, at $T=1$ we have
still not reached thermal equilibrium within our time scale.
Once this is understood, we proceed to extrapolate to infinite time as
\begin{equation}
\frac{G_\text{L}(r,t_\text{tot})}{G_\text{R}(r,t_\text{tot})}=
\frac{G_\text{L}(r,t_\text{tot}=\infty)}{G_\text{R}(r,t_\text{tot}=\infty)}+
\frac{A(r)}{t_\text{tot}^x}\,.
\end{equation}
In the above equation, the exponent $x$ was allowed to depend on $r$ and $T$ (we found
that it barely depended on $r$ for a given temperature). We were able to carry out
this extrapolation safely down to temperature $T=1.0$, see right panel in
Fig.~\ref{fig:GL_GR2}. In the limit of long times, the ratio of propagators
does decrease exponentially with $r$, which confirms that
$\xi_\text{R}>\xi_\text{L}$ (at least down to temperature $T=1.0$, for
$H=0.1$ and $H=0.2$, and assuming that the algebraic prefactors in the
ratio $G_L/G_R$ are not relevant).

\end{document}